\documentstyle[preprint,aps]{revtex}

\newcommand{\be}{\begin{equation}}
\newcommand{\ee}{\end{equation}}
\newcommand{\bea}{\begin{eqnarray}}
\newcommand{\eea}{\end{eqnarray}}
\newcommand{\non}{\nonumber}

\newtheorem{th}{Theorem}[section]
\newtheorem{lem}{Lemma}[section]

\newtheorem{prop}{Proposition}[section]
\newcommand{\ra}{\rangle}

\newcommand{\lam}{\lambda} 
 
\newcommand{\al}{\alpha}

\begin{document}
\title{
Non-regular eigenstate of the XXX model 
as some limit of the Bethe state 
}
\author{ Tetsuo Deguchi}
\maketitle
\begin{center} 
Department of Physics, Faculty of Science \\
      Ochanomizu University \\
         2-1-1 Ohtsuka, Bunkyo-Ku, Tokyo 112-8610, Japan .   
\footnotetext[1]{deguchi@phys.ocha.ac.jp .} 
\end{center} 

\begin{abstract}
For the one-dimensional XXX model under the periodic boundary conditions,
we discuss two types of eigenvectors, 
regular eigenvectors  which have finite-valued rapidities 
satisfying the Bethe ansatz equations, 
and non-regular eigenvectors  which are descendants 
of some regular eigenvectors 
under the action of the SU(2) spin-lowering operator. 
It was pointed out by many authors that the non-regular eigenvectors 
should correspond to the Bethe ansatz wavefunctions which have multiple  
infinite rapidities.  However, it has not been explicitly 
shown whether such a delicate limiting procedure should be possible.   
In this paper, we discuss it explicitly 
in the level of  wavefunctions: we prove 
that any non-regular eigenvector of the XXX model 
is derived  from the Bethe ansatz wavefunctions 
through some limit of infinite rapidities. 
We formulate the regularization also in terms of 
the algebraic Bethe ansatz method. 
As an application of infinite rapidity, 
we discuss the period of the spectral 
flow under the twisted periodic boundary conditions. 
\end{abstract} 
 %

 \newpage
  \setcounter{equation}{0} 
  \renewcommand{\theequation}{1.\arabic{equation}}
 \section{Introduction} 
 The one-dimensional Heisenberg model (XXX model) under the periodic 
 boundary conditions is one of the fundamental models of integrable  
 quantum spin systems. \cite{Bethe} 
 Under the spin $SU(2)$ symmetry 
 any eigenvector of the Hamiltonian is given by a highest-weight vector 
 or a descendant of some highest-weight vector. It has been shown 
 by  the algebraic Bethe ansatz method  \cite{algebraicBA}  
 that any regular Bethe ansatz eigenstate of 
 the XXX model  gives  a highest weight vector. \cite{TF,TF-reviews}.   
 Let us consider the XXX Hamiltonian  under the periodic boundary conditions
  \be 
  {\cal H} = -{\frac J 4} \sum_{\ell=1}^{L} {\vec \sigma }_{\ell} \cdot 
  {\vec \sigma}_{\ell +1}, \quad {\rm where} \quad   
  {\vec \sigma}_{L+1}={\vec \sigma}_1 \, . 
  \ee 
  Here the symbol ${\vec \sigma}_{\ell}= 
  (\sigma_{\ell}^x, \sigma_{\ell}^y, \sigma_{\ell}^z) $ 
  denotes the spin angular-momentum operator 
  with $S=1/2$ acting on the $\ell$-th site of the ring. 
  Let us denote by the symbol  ${\vec S}_{tot}$ the total spin-operator:  
  ${\vec S}_{tot}= \sum_{\ell=1}^{L} {\vec \sigma}_{\ell}/2$.  
 Then, it is easy to show that the Hamiltonian 
 is invariant under the action of 
  the $SU(2)$:    $ [{\cal H}, {\vec S}_{tot}]=0$.

\par  
 Let us introduce some notation   
  of the coordinate Bethe ansatz  
 \cite{Bethe,Yang-Yang,Yang}.      
  We denote by $x_1, x_2, \ldots$, $x_M$     
  the coordinates of the $M$ down-spins set  
  in increasing order:  $1 \le x_1 < x_2 < \cdots < x_M \le L$.   
  Then, we  define the Bethe ansatz  wavefunction with $M$ parameters $k_1$, $k_2$,  
   $\dots$, $k_M$  by the following: 
  \be
  f_M^{(B)}(x_1,\ldots, x_M; k_1, \ldots, k_M) =\sum_{P \in {\cal S}_M} A_M(P) 
  \exp \left( i\sum_{j=1}^M k_{Pj} x_j \right) ,  
  \label{BAWF}
  \ee 
  where the sum is over all the permutations of $M$ letters of the set 
  $\{1, 2, \ldots, M \}$,  
  and the symbol $Pj$ denotes the action of permutation 
  $P$ on letter $j$ .  
  Here the symbol ${\cal S}_M $ denotes the permutation group of $M$ letters. 
  We define the amplitudes $A_M(P)$'s of the Bethe ansatz 
  wavefunction  by   
  \be
  A_M(P) = 
   C \, \epsilon(P) \prod_{1 \le j < \ell \le M} 
  {\frac 
   {\exp[i(k_{Pj}+k_{P{\ell}})] +1 -2 \exp(ik_{Pj}) }
   {\exp[i(k_j+k_{\ell})] +1 - 2 \exp(ik_j) } } ,  
  \quad {\rm for } \quad P \in {\cal S}_M . 
  \label{AM1}
  \ee 
 Here the symbol $\epsilon(P)$ denotes the sign of  permutation $P$, 
 and $C$ is a constant.  Let the symbol $|0 \ra $ denote the vacuum state 
  where all spins are up ($M=0$).  Then, 
 we construct the following vector from  the Bethe ansatz wavefunction   
  \be
  || M \ra = \sum_{1 \le x_1< x_2 < \cdots < x_M \le L} 
  f_M^{(B)}(x_1, \ldots, x_M; k_1, \ldots, k_M) 
  \sigma_{x_1}^{-} \sigma_{x_2}^- \ldots \sigma_{x_M}^- |0\ra.   
  \label{FBV}
  \ee
   Here,  the summation is over all the possible values of $x_j$'s 
  given in increasing order. 
   We call the vector $|| M \ra $ (\ref{FBV}) 
  with the amplitudes defined by eq. (\ref{AM1}), 
  a {\it formal Bethe vector} (or 
    {\it formal Bethe state}). 
  We recall that there is no constraint on 
   the $M$ parameters  $k_1, k_2, \ldots$, $k_M$.  When they are generic,  
  the formal Bethe state 
  (\ref{FBV}) is not  an eigenvector of the XXX Hamiltonian.

 \par 
 Now, let us consider the Bethe ansatz equations. They correspond to   
 the periodic boundary conditions for the Bethe ansatz wavefunction.    
 \bea
  \exp(i L k_j ) & = &
  (-1)^{M-1} \prod_{\ell=1, \ell \ne j}^{M} 
  {\frac 
   {\exp[i(k_{j}+k_{\ell})] +1 -2 \exp(ik_{j}) }
   {\exp[i(k_j+k_{\ell})] +1 - 2 \exp(ik_{\ell})}  } , \non \\
  & & \qquad  
  {\rm for} \quad j=1,\ldots, M.
  \label{BAEm}
  \eea
 If all the parameters $k_1, k_2, \ldots$,  $k_M$ satisfy the Bethe ansatz 
 equations,  then the formal Bethe vector $|| M \ra$ becomes an eigenvector 
 of the XXX Hamiltonian. Furthermore, if the $k_j$'s 
 satisfy the conditions that 
 $k_j \ne 0 \, ({\rm mod} \, 2\pi)$ for $j=1, \ldots, M$, then  
 we call the eigenvector  {\it regular}, and denote it by the symbol 
 $| M \ra$. It is called regular, since it is well defined as 
 an eigenstate given by the Bethe ansatz wavefunction. 
 In this sense, it is also called a regular Bethe ansatz state or 
 a Bethe state, in short.

 \par 
 A regular eigenstate can lead to a series of    
 non-highest weight eigenvectors of the $SU(2)$ symmetry. 
  Let $| R \ra$ denote a given regular eigenstate 
  with $R$ down-spins. Then, it is a highest weight vector 
   of the $SU(2)$ symmetry with $S_{tot}=L/2 - R$ and 
   $S^z_{tot}=L/2-R$.   
 Here we assume that the number $R$ should satisfy the condition:  
 $0 \le R \le L/2$, for regular eigenvectors.  
  From the eigenvector $| R \ra$, 
  we can derive a sequence of non-highest weight eigenvectors: 
  $(S_{tot}^{-})^K |R\ra$ for  $K=1, \ldots, L-2R$. 
 We call the series of  descendant eigenstates 
 {\it  non-regular}. We denote them by 
 \be 
 |R, K \ra = {\frac 1 {K!}} \left( S_{tot}^{-}  \right)^{K} |R\ra 
 \quad {\rm for} \quad  K=1, \ldots, L-2R. 
 \label{RK}
 \ee
 It is remarked that the eigenvectors $| R, K \ra$'s are fundamental 
  in the completeness of the spectrum of the XXX model,  
 although they are called non-regular in this paper. 

 \par 
 The main question of this paper is 
 how non-regular eigenvectors 
 of the XXX model are related to the Bethe ansatz wavefunctions.   
 In fact, it has already been observed by Gaudin \cite{Gaudin} that 
 the non-regular eigenvectors  are associated with 
 the Bethe ansatz wavefunction with several parameters 
 $k_j$'s being equal to zero. 
  Furthermore, it was shown by Takhtajan and Faddeev \cite{TF} that  
 the creation operator $B(v)$ is equivalent to the spin-lowering 
 operator $S^{-}_{tot}$  by sending the rapidity $v$ to infinity 
 (see also Refs. \cite{Woynarovich,Essler1,Sutherland,PR}). 
 We note that the limit of sending the parameter $k_j$ to zero corresponds to the limit of 
 the infinite rapidity. Here, for a given parameter $k$, the rapidity $v$ has been defined   
 by the relation: $\exp(i k) = (v + i )/(v - i)$;  rapidity $v$ is finite   
 if and only if $k \ne 0$ (mod $2 \pi$).  In spite of the observations, 
 however, it has not been clearly shown yet  whether one can 
 construct the non-regular eigenvector $|R, K \ra$ from the Bethe ansatz wavefunctions 
 for the  case of general $K$.  In the case of multiple infinite rapidities,  
 the limit of the wavefunction depends  not only on its normalization  but also 
 on how we control  the differences among the infinite rapidities.  
 Thus, under a naive limiting procedure, 
 the amplitudes of the formal Bethe state become indefinite; it can vanish or diverge 
 depending on the limiting procedure. 
  (For example, see also \cite{Siddharthan}.)  
  Furthermore, if a set of parameters $k_j$'s contains  multiple zeros, 
 then  it is not clear whether the Bethe ansatz wavefunction should vanish or not. 
 In fact, for any given regular eigenvector, we can show that  
 if  two momenta (or two rapidities) have the same value, 
 then the norm of the eigenvector is given by zero. This fact is  called   
 the ``Pauli principle'' of the Bethe ansatz wavefunction.  
 Thus, the question  has  been nontrivial. 
 In this  paper, we make it clear.  
 We show that there exists a certain limiting procedure through which  
  any  non-regular eigenvector of the XXX model is derived  
 from the formal Bethe state.

 \par 
  Let us explain our derivation of non-regular eigenvectors 
from the formal Bethe states, briefly. 
 We consider  a given regular Bethe ansatz eigenstate $| R \ra$ 
 with $R$ down-spins.  It has  $R$ rapidities 
 $v_1, v_2, \ldots, v_R$,   satisfying  
 the Bethe ansatz equations for $R$ down-spins.  
 For a given positive integer $K$, we consider the non-regular eigenstate 
 $|R,K \ra $. We recall that it has been  defined in eq. (\ref{RK}) and 
 is derived from $|R \ra$. 
  Then, we introduce an additional set of the rapidities 
   $v_{R+1}$, $\ldots$, $v_{R+K}$ as follows 
 \be
 v_{R+j}(\Lambda) = \Lambda  + \delta_{j}, \quad {\rm for} 
  \quad j=1, \ldots, K \, . 
 \label{Kset}
 \ee
 Here we call the parameter $\Lambda$ the ``center'' of 
 the additional $K$ rapidities $v_{R+1}, \ldots, v_{R+K}$.  
 We assume that the $\delta_{j}$ 's are arbitrary non-zero parameters, 
 which can be sent to infinity.  
  Let us now consider a formal Bethe vector 
  $||R+K\ra$ with $R+K$ down-spins  
   that has $R$ rapidities 
   of the given regular eigenstate $|R\ra$ ({\it i.e.}, $v_1, \ldots, v_R$)
   together with the additional $K$ rapidities given by eq. (\ref{Kset}) 
({\it i.e.}, $v_{R+1}(\Lambda), \ldots, v_{R+K}(\Lambda)$).  
   We denote it by $||R, K; \Lambda \ra$. 
  Then,  we can show that 
  the vector $||R,K; \Lambda\ra$  becomes the non-regular eigenstate   
   $|R,K\ra$ by sending $\Lambda$ to infinity:  
 \be
 \lim_{\Lambda \rightarrow \infty} ||R,K; \Lambda\ra = C \, |R,K \ra 
 \ee
 Here $C$ denotes a constant. 
  Thus,   the  non-regular eigenstate is derived 
    from the Bethe ansatz wavefunction.

 \par 
  We discuss only regular eigenvectors of the XXX model 
 and  their descendants which we call non-regular eigenvectors. 
 We do not consider  other types of solutions  in this paper. In fact, 
 it was shown that the so-called string hypothesis predicts the correct number 
 of appropriate solutions to the Bethe ansatz equations of the 
 XXX model under the periodic boundary conditions  
  \cite{Bethe,Takahashi,Kirillov}. 
 Although the hypothesis fails to count the particular type of solutions, 
 all the known numerical or analytical researches have shown that 
 the total number of solutions to the Bethe ansatz equations 
 is given correctly \cite{Bethe,EKS,Essler2,PR}. 
 Thus, it is conjectured  that all the regular eigenvectors and  their descendants 
 give the complete set of eigenvectors of the XXX model. 
 In fact, it is  proven that the number of solutions of the Bethe ansatz equations 
 is given correctly  for the XXX model under the  
 twisted boundary conditions with the {\it generic} twisting parameter 
  \cite{Tarasov-Varchenko}.   
 It seems that the theorem does not cover the case of the periodic boundary conditions,  
 since it  corresponds to a non-generic point of the twisting parameter.   
 However, the result of the paper 
 might also shed some light on the mathematical understanding  
 of the string hypothesis and the number counting arguments in general, 
  as we shall discuss in sections 5 and 6.  

 \par 
  The contents of the paper consists of the following. 
  In section 2 we give a formula describing  
  the action of powers of the spin-lowering operator.   
  Then, through some examples, we explicitly discuss the derivation of  
 non-regular eigenvectors from  formal Bethe states. 
  It is shown that  infinite rapidities do not always satisfy the Bethe ansatz equations,  
 although the limit of the Bethe ansatz wavefunction  satisfies  
 the periodic boundary conditions.  
 In section 3, we give an explicit proof for the construction 
  of  non-regular eigenstates from the  formal Bethe states.  
 We show in section 4 that  the formal Bethe state can be defined naturally 
  in the algebraic Bethe ansatz. 
 In fact,  the formal Bethe state $|| M \ra $ is 
 equivalent to the vector generated by the $B$ operators on the vacuum:  
 $B(v_1) \cdots B(v_M) |0 \ra$  with the $M$ rapidities $v_1, \ldots, v_M$ being generic.    
 In section 5, we show how the concept of formal Bethe states is useful in the analysis of 
  the spectral flow of the XXX model under the twisted boundary conditions.  
 In fact, we can derive 
 the $4\pi$-period of the spectral flow under the twisted boundary conditions, 
  almost rigorously. In section 6,  we give some discussions. 
In order to make the paper self-consistent, some Appendices are provided. 
 The formula for the action of spin-lowering operator 
 is proven in Appendix A.   
An example of the formal Bethe state with three infinite rapidities 
is discussed  in Appendix B. 
Some fundamental properties of the symmetric group are given in Appendix C,  
which are important in sec. 3. The ``Pauli principle'' 
of the Bethe ansatz wavefunction is explicitly proven  in Appendix D. 
Finally, we formulate rigorously 
the coordinate Bethe ansatz method introduced by Bethe \cite{Bethe} 
in Appendix E.


   \setcounter{equation}{0} 
   \renewcommand{\theequation}{2.\arabic{equation}}
 \section{Formal Bethe states and non-regular eigenstates }

  \subsection{Non-regular eigenstates}

\par 
Let us discuss the action of spin-lowering operator 
on arbitrary vectors with $M$ down-spins, explicitly. 
For an illustration we consider  the case of $M=1$. 
Let $|1)$ denote a  vector with one down-spin 
\be
|1)= \sum_{x_1=1}^{L} g(x_1) \sigma_{x_1}^{-} |0 \ra \, ,  
\ee
where  $g(x)$ is any given arbitrary function. 
 Applying  to it the spin-lowering operator 
$S_{tot}^{-} = \sum_{j=1}^{L} \sigma_{j}^{-}$, 
we have   
\bea 
 S_{tot}^{-} |1) & = &\sum_{x_2=1}^{L} \sigma_{x_2}^{-} 
 \sum_{x_1}^{L} g(x_1) \sigma_{x_1}^{-}|0 \ra  \non \\
& =& \sum_{x_1=1}^{L} \sum_{x_2=1}^{L} 
g(x_1) 
\sigma_{x_1}^{-}\sigma_{x_2}^{-} |0 \ra   \non \\
& = & \left( \sum_{1 \le x_1< x_2 \le L} +\sum_{1 \le x_2 < x_1 \le L} \right)  
g(x_1) 
\sigma_{x_1}^{-}\sigma_{x_2}^{-} |0 \ra   \non \\
&=& \sum_{1 \le x_1 < x_2 \le L} \left( g(x_1) +g(x_2) 
\right) \sigma_{x_1}^{-} \sigma_{x_2}^{-} |0 \ra  
\non \\
&=& \sum_{1 \le x_1 < x_2 \le L} \left(\sum_{1 \le j \le 2} g(x_j) 
\right) \sigma_{x_1}^{-} \sigma_{x_2}^{-} |0 \ra  
\label{M=1}
\eea
Here we note that $\left( \sigma_{x}^{-} \right)^2 | 0 \ra  =0$.

 \par 
 We can generalize the expression (\ref{M=1}). 
 Let us denote by the symbol $|M)$ a vector with  $M$ down-spins 
 \be 
|M)= \sum_{1 \le x_1 < x_2 < \cdots < x_M \le L} 
g(x_1, x_2, \cdots, x_M) \,
 \sigma_{x_1}^{-}  \sigma_{x_2}^{-} \cdots \sigma_{x_M}^{-} \, |0 \ra  \, ,  
\label{|M)}
\ee 
 where  $g(x_1,x_2, \cdots, x_M)$ is     
 an arbitrary function of $x_j$'s.  Then,   
 it is clear that any vector with $M$ down-spins can be 
 considered  as a vector $|M)$ with some function 
 $g(x_1, x_2, \cdots, x_M)$. 
 Now, we introduce  the following formula  
  \be
  {\frac 1 {K!}}  \left( S_{tot}^{-} \right)^K |M) 
 = \sum_{1 \le x_1< \cdots < x_{M+K} \le L} 
   \left( \sum_{1 \le j_1 <  \cdots < j_M \le M+K}
    g(x_{j_1}, \ldots, x_{j_{M}}) \right) 
   \sigma_{x_1}^{-} \cdots \sigma_{x_{M+K}}^{-} |0 \ra  .  
   \label{vec}
   \ee
 We note that the expression (\ref{M=1}) corresponds to the 
 case  $M=K=1$.  
  An explicit  proof of the formula (\ref{vec}) will be given in Appendix A. 
  In sec. 2.C, we shall consider the special case 
 of $K=2$ and $M=1$, which is given in the following      
 \bea 
 {\frac 1 2}
 \left(S_{tot}^{-}  \right)^2 |1)  
&  = & \sum_{1 \le x_1 <x_2 <x_3 \le L} \sum_{1 \le j \le 3} g(x_j) 
 \sigma_{x_1}^{-} \sigma_{x_2}^{-} \sigma_{x_3}^{-} |0 \ra \non \\
&  = & \sum_{1 \le x_1<x_2<x_3 \le L} \left( g(x_1) + g(x_2) + g(x_3) \right) 
 \sigma_{x_1}^{-} \sigma_{x_2}^{-} \sigma_{x_3}^{-} |0 \ra  .
 \label{lower2}
 \eea
 Here we note that  $M+K=1+2=3$.

 \par 
  Let us consider a regular eigenstate $|R \ra $ 
   with $R$ down-spins, and  the  non-regular eigenstate $|R,K \ra $ 
   given by eq. (\ref{RK}).   
   We recall that $|R \ra $ 
   is a highest weight vector
   of the $SU(2)$  with $S=L/2-R$ and $S_z=L/2-R$.  
  By applying the formula (\ref{vec}) to the definition 
  (\ref{RK}) of the non-regular eigenvector, 
   then it  is explicitly expressed 
   in terms of the Bethe ansatz wavefunctions  
   \be
   |R,K \ra  = \sum_{1 \le x_1< \cdots < x_{R+K} \le L} 
   \left( \sum_{1 \le j_1 <  \cdots < j_R \le R+K}
    f_{R}^{(B)}(x_{j_1},\ldots,x_{j_{R}}) \right) 
   \sigma_{x_1}^{-} \cdots \sigma_{x_{R+K}}^{-} |0 \ra  .  
   \label{vecRK}
   \ee
 Here we recall that the function 
$f_R^{(B)}(x_1, \cdots x_R; k_1, \ldots, k_M)$ 
is the  Bethe ansatz wavefunction defined in eq. (\ref{BAWF}), where 
the $k_j$ 's satisfy the Bethe ansatz equations.

 \subsection{Amplitudes of formal Bethe states} 

Let us recall the relation between  rapidity $v_j$ and parameter $k_j$ 
\be 
\exp(ik_j) ={ \frac {v_j +i} {v_j -i}}
\quad {\rm for} \quad j=1, \ldots, M . 
\label{rapidity}
\ee
In terms of rapidities,  the  Bethe ansatz equations are given by   
   \be
   \left( {\frac {v_j+i} {v_j-i}} \right)^L = 
   \prod^{M}_{\ell=1,  \ell \ne j} \left( 
   {\frac {v_j - v_{\ell} + 2i } 
   {v_j - v_{\ell} - 2i} } \right) , 
   \qquad  {\rm for} \quad j=1,\ldots, M. 
   \label{BAEr} 
   \ee
The amplitudes $A_M(P)$'s  defined in eq. (\ref{AM1}) are given by the following 
\be 
A_M(P)\left[ v_1, \cdots, v_M \right] 
 = \epsilon(P) \prod_{1 \le j < k \le M} 
 {\frac {v_{Pj}- v_{Pk} + 2 i} {v_{j}- v_{k} + 2 i} } 
\label{amp0}
\ee
Here, the dependence of the amplitude $A_M(P)$ on 
rapidities $v_1, \ldots, v_M$ is expressed 
in the bracket $[ \cdots ]$, explicitly.  Here we note that 
the expression (\ref{AM1}) of the amplitude $A_M(P)$ is proven 
in Appendix E. 

\par 
 Let us now introduce a useful formula for expressing the amplitudes 
 of the Bethe ansatz wavefunction. 
 We denote by   the symbol $H(x)$  the Heaviside step function 
  defined by $H(x) = 1$ for $x>0$, and $H(x) =0$ otherwise.   
 Then, we can show that  
  the amplitudes $A_M(P)$'s given in eq.(\ref{amp0})  are expressed 
   by the following 
\be 
A_M(P) =  \prod_{1 \le j < k \le M}
  \left( {\frac {v_j-v_k -2i}  {v_j-v_k + 2i}}  
 \right)^{H(P^{-1}j - P^{-1}k)}   
\label{amp}
\ee
We shall prove the expression (\ref{amp}) in sec. 3.

\par 
  For an illustration, we consider the amplitudes $A_M(P)$'s for the case $M=3$.  
 Let us express $A_M(P)$ by $A_{P1P2\cdots PM}$. Then, they 
 are given as follows  
\bea
A_{123}& = & 1, \quad 
A_{132}= {\frac {v_2 -v_3 -2 i } {v_2-v_3 +2i}}, \quad 
A_{213}= {\frac {v_1 -v_2 -2 i } {v_1-v_2 +2i}}
,  \non \\
A_{231} & =& \left( 
{\frac {v_1 -v_2 -2 i } {v_1-v_2 +2i}}
\right) 
\left( 
{\frac {v_1 -v_3 -2 i } {v_1-v_3 +2i}} \right) , \non \\ 
A_{312} & = &
\left( 
{\frac {v_1 -v_3 -2 i } {v_1-v_3 +2i}} \right) 
\left( 
{\frac {v_2 -v_3 -2 i } {v_2-v_3 +2i}} \right) , \non \\
A_{321}& = & 
\left( 
{\frac {v_1 -v_2 -2 i } {v_1-v_2 +2i}} \right) 
\left( 
{\frac {v_1 -v_3 -2 i } {v_1-v_3 +2i}} \right) 
\left( 
{\frac {v_2 -v_3 -2 i } {v_2-v_3 +2i}} \right) . 
\label{samples}
\eea

\subsection{Formal Bethe states with additional infinite rapidities}

Let us discuss  some examples of the Bethe ansatz 
wavefunctions with additional rapidities. 
We first consider the case of  
  three down-spins with $R=1$ and $K=2$, 
{\it i.e.}, the formal Bethe state $||1,2;\Lambda \ra $. 
Here, $v_2$ and $v_3$ are  additional rapidities defined 
by eq. (\ref{Kset}): $v_2=\Lambda + \delta_1 $, 
$v_3= \Lambda +\delta_2$.  Here we assume that $\delta_1$ and $\delta_2$ 
are some constants.  
We recall that $v_1$ is the rapidity of 
the state $|1 \ra$ and it satisfies 
the Bethe ansatz equation for $M=1$.

\par 
 Let us denote the difference $\delta_1-\delta_2$ by $ \Delta$.
 For simplicity, we assume that $\delta_1=-\delta_2$. Then, the 
additional rapidities are given by 
$v_2= \Lambda + \Delta/2$ and $v_3=\Lambda- \Delta/2$.    
Substituting the rapidities $v_1$, $v_2$ and $v_3$ 
into the amplitudes in  (\ref{samples}), 
we have
\bea
A_{123}(\Lambda)& = & 1, \quad 
A_{132}(\Lambda)= {\frac {\Delta -2 i } {\Delta +2i}}, \quad 
A_{213}(\Lambda)= {\frac {v_1 -\Lambda - \Delta/2 -2 i } 
         {v_1-\Lambda - \Delta/2 +2i}},  \non \\
A_{231}(\Lambda) & =& \left( 
{\frac {v_1 -\Lambda -\Delta/2 -2 i } {v_1- \Lambda - \Delta/2 +2i}}
\right) 
\left( 
{\frac {v_1 - \Lambda+ \Delta/2 -2 i }
 {v_1-\Lambda + \Delta/2 +2i}} \right) , \non \\ 
A_{312}(\Lambda) & = &
\left( 
{\frac {v_1 - \Lambda+ \Delta/2 -2 i } {v_1- \Lambda+ \Delta/2 +2i}} \right) 
\left( 
{\frac {\Delta -2 i } {\Delta +2i}} \right) , \non \\
A_{321}(\Lambda)& = & 
\left( 
{\frac {v_1 -\Lambda- \Delta/2 -2 i } {v_1-\Lambda- \Delta/2 +2i}} \right) 
\left( 
{\frac {v_1 -\Lambda+ \Delta/2 -2 i } {v_1-\Lambda+ \Delta/2 +2i}} \right) 
\left( 
{\frac {\Delta -2 i } {\Delta +2i}} \right) . 
\label{samples2}
\eea
Let us denote by $f_{R,K}^{(B)}$ the Bethe ansatz wavefunction for 
the formal state $||R,K; \Lambda \ra$. 
The Bethe ansatz wavefunction of $||1,2; \Lambda \ra $ 
is given by 
\bea
& & f_{1,2}^{(B)}(x_1,x_2,x_3;k_1,k_2(\Lambda), k_3(\Lambda))  
\non \\
& =& 
A_{123} 
\exp { i \left(k_1x_1+k_2(\Lambda) x_2 + k_3(\Lambda) x_3 \right)} 
 +A_{132} 
\exp { i \left( k_1 x_1+k_3(\Lambda) x_2 + k_2(\Lambda) x_3 \right)} 
\non \\
& + & A_{213}  
\exp{i \left(k_2(\Lambda)x_1+  k_1 x_2 + k_3(\Lambda) x_3 \right)} 
+ A_{312}
\exp{i \left(k_3(\Lambda)x_1+  k_1 x_2 + k_2(\Lambda) x_3 \right)} 
\non \\
&+ & A_{231} \exp { i \left(k_2(\Lambda) x_1+k_3(\Lambda) x_2 + k_1 x_3 \right)} 
+A_{321} \exp { i \left(k_3(\Lambda) x_1+k_2(\Lambda) x_2 + k_1 x_3 \right)} \non \\
& & \qquad  {\rm for} \quad 1 \le x_1< x_2 < x_3 \le L \, , 
\eea
where $k_2(\Lambda)$ and $k_3(\Lambda)$ are given by  
\be
\exp(ik_2(\Lambda)) = \left( {\frac {\Lambda +\Delta/2 + i}
 {\Lambda +\Delta/2 -i }} \right)\, , \quad 
\exp(ik_3(\Lambda)) = \left( {\frac {\Lambda -\Delta/2 + i}
 {\Lambda - \Delta/2 -i }} \right) \, . 
\ee
Sending the center $\Lambda$ to infinity : $\Lambda \rightarrow \infty$, 
we have  $k_2=k_3=0 \, ({\rm mod}) 2\pi$ and  
\bea 
A_{123}(\infty)& =& A_{213}(\infty) = A_{231}(\infty) = 1 \non \\
A_{132}(\infty) & = & A_{312}(\infty) = A_{321}(\infty) = 
{\frac {\Delta - 2i}{ \Delta + 2i}} \, . 
\eea
Therefore, the limit of the Bethe ansatz wavefunction is given by 
\be 
\lim_{\Lambda \rightarrow \infty}
 f_{1,2}^{(B)}(x_1,x_2,x_3;k_1,k_2(\Lambda), k_3(\Lambda)) = C_2 \, 
\left( e^{ik_1 x_1} + e^{ik_1 x_2} + e^{ik_1x_3} \right) \, .
\label{limit}
\ee
where the constant $C_2$ is given by 
\be 
C_2 = \left(1 + {\frac {\Delta -2i}{\Delta +2i}} \right) 
\ee
Combining the eqs. (\ref{limit}) and (\ref{lower2}), 
we obtain the following result.  
\bea
 \lim_{\Lambda \rightarrow \infty} ||1,2;\Lambda \ra  
& = & C_2 \, 
 \sum_{1 \le x_1<x_2<x_3 \le L} \left(
e^{ik_1x_1} + e^{ik_1 x_2} + e^{ik_1 x_3} \right) 
\sigma_{x_1}^- \sigma_{x_2}^{-} \sigma_{x_3}^{-} |0 \ra  \non \\
& = & C_2 \, 
 {\frac 1 {2!}} \left( S_{tot}^{-} \right)^2 |1 \ra  \non \\
& = & C_2 \, |1,2 \ra 
\label{res}
\eea
Thus, we have shown that the limit of the formal Bethe state $|| 1,2; \Lambda \ra $ 
is equivalent to the non-regular eigenstate $|1,2 \ra $. 
We shall prove this equivalence  for the general case in sec. 3. 
For an illustration, we shall consider the case of $R=0$ and $K=3$ in Appendix B. 

\par 
 Let us give some remarks on  eq. (\ref{res}). 
 We see that the limiting procedure depends 
 on  the difference $\Delta$. If $\Delta=- 2 i$, then the constant $C_2$ becomes infinite. 
If $\Delta=0$, then the constant $C_2$ vanishes.  Thus, the limit of the  wavefunction  
with infinite rapidities $v_2$ and $v_3$ 
 depends on how we send them  into infinity.

\subsection{The P.B.C.s for the limits of the formal Bethe states }

The formal Bethe state $|| R,K; \Lambda \ra $ 
satisfies the periodic boundary conditions after taking the limit: 
$\Lambda \rightarrow \infty$.  In fact, 
it is clear since the limit gives the 
non-regular eigenvector $|R,K \ra $, 
which satisfies the periodic boundary conditions. 
Here we note that the total spin operator 
${\vec S}_{tot}$ is translation  invariant. 
However,  infinite rapidities do not always  satisfy the 
Bethe ansatz equations. 

\par 
For an illustration,   let us consider the formal Bethe state $||1,2; \Lambda \ra$.  
We denote by $f_{1,2}^{(\infty)}(x_1,x_2,x_3)$  
the limit of $f_{1,2}^{(B)}(x_1,x_2,x_3;k_1,k_2(\Lambda), k_3(\Lambda))$ 
 with $\Lambda$ sent to infinity.  
We see that it satisfies the periodic boundary conditions: 
$f_{1,2}^{(\infty)}(x_1,x_2,x_3) = f_{1,2}^{(\infty)}(x_2,x_3,x_1+L)$ for  
$1 \le x_1 < x_2 < x_3 \le L$.  Explicitly we have  
\be 
f_{1,2}^{(\infty)}(x_2,x_3,x_1+L)= 
{\frac {2 \Delta}{\Delta+ 2i}}
\left(e^{ik_1 x_2} + e^{ik_1 x_3} + e^{ik_1 (x_1 + L)} \right)\, . 
\ee
Thus, it satisfies the periodic boundary conditions if and only if 
the following holds:
\be 
\exp(ik_1 L) = 1 
\ee
This is nothing but the Bethe ansatz equation for $k_1$, 
and it does hold from the assumption that $v_1$ is the rapidity of a 
regular eigenvector $|1 \ra$.

\par 
Now let us show  that  the additional rapidities do not necessarily 
satisfy the Bethe ansatz equations, 
although the limiting Bethe ansatz wavefunction 
satisfies the periodic boundary conditions. 
Let us consider the Bethe ansatz equations for 
three rapidities $v_1, v_2$ and $v_3$
\bea 
\left( {\frac {v_1 + i } {v_1 -i}} \right)^{L}  
& = & 
\left( {\frac {v_1 - v_2 + 2i } {v_1 - v_2 - 2i}} \right)  
\left( {\frac {v_1 - v_3 + 2i } {v_1 - v_3 - 2i}} \right)  
\non \\
\left( {\frac {v_2 + i } {v_2 -i}} \right)^{L}  
& = & 
\left( {\frac {v_2 - v_1 + 2i } {v_2 - v_1 - 2i}} \right)  
\left( {\frac {v_2 - v_3 + 2i } {v_2 - v_3 - 2i}} \right)  
\non \\
\left( {\frac {v_3 + i } {v_3 -i}} \right)^{L}  
& = & 
\left( {\frac {v_3 - v_1 + 2i } {v_3 - v_1 - 2i}} \right)  
\left( {\frac {v_3 - v_2 + 2i } {v_3 - v_2 - 2i}} \right)  
\label{BA3}
\eea
Taking the limit: $\Lambda \rightarrow \infty$, the three equations are 
reduced into the following  
\bea 
\left( {\frac {v_1 + i } {v_1 -i}} \right)^{L}  
& = & 1 \label{LBA1} \\
\left( {\frac {\Delta_{} + 2i } {\Delta_{} - 2i}} \right)  
& = & 1 
\label{LBA2}   
\eea
The equation (\ref{LBA2}) does not hold if $\Delta_{}$ takes a finite value;  
it holds only if $|\Delta|= \infty$.


   \setcounter{equation}{0} 
   \renewcommand{\theequation}{3.\arabic{equation}}
  \section{Proof of the limit of formal Bethe states}

  In this section we prove the theorem in the following. 
  \begin{th} 
  Let $|R \ra$ be a regular Bethe ansatz eigenstate with $R$ down-spins 
  and rapidities $v_1, \ldots, v_R$. 
  We recall that the symbol $||R,K; \Lambda \ra$ denotes 
  the formal Bethe state with $R+K$ down-spins, 
 which has the $R$ rapidities $v_1$, \ldots, $v_R$ of $|R \ra$ together with 
 additional rapidities $v_{R+1}(\Lambda) \ldots, v_{R+K}(\Lambda)$. 
  Then,   the non-regular eigenstate $|R,K \ra$, which is a descendant of $R$,  
  is equivalent to  the limit of the formal Bethe state 
  $||R, K; \Lambda \ra$  with $\Lambda$ sent to infinity:    
  \be 
  \lim_{\Lambda \rightarrow \infty} ||R, K; \Lambda \ra = C_K |R,K \ra  
  \label{lim}
  \ee
 \end{th}

\subsection{Derivation of the formula for amplitudes $A_M(P)$'s }

We now discuss the derivation of the formula (\ref{amp}), 
which rewrites the amplitudes $A_M(P)$'s 
 defined in (\ref{AM1}). 
Let us recall that 
the symbol $H(x)$ denote the Heaviside step function 
defined by $H(x) = 1$ for $x>0$, and $H(x) =0$ otherwise.   
We now show the following proposition. 
\begin{lem} 
Let $P$ be an element of  ${\cal S}_M$, 
and $v_1, v_2, \ldots, v_M$ be generic parameters.  
Then, the following identity holds. 
\be
\prod_{1 \le j<k \le M} 
{\frac {v_{Pj}-v_{Pk} + 2i } {v_{j}-v_{k} + 2i } }
 = \prod_{1 \le j<k \le M} \left( 
{\frac {v_{k}-v_{j} + 2i }  {v_{j}-v_{k} + 2i } }
\right)^{H(P^{-1}j-P^{-1}k)} .  
\label{A1}
\ee
\end{lem}
({\it Proof})
Let us take  a pair of integers $j$ and $k$ with $j<k$, 
 and consider the factor $v_j-v_k + 2i$ in the denominator of 
LHS of eq. (\ref{A1}). For the pair, 
there exist two integers 
$\ell$ and $m$ such that $P \ell =j$, $Pm=k$. 
There are two cases either $\ell < m$ or $\ell > m$. 
If $\ell < m$,  then we have the factor 
${v_{P\ell} -v_{Pm}+ 2i }$ in the enumerator of LHS of eq. (\ref{A1}). 
Thus,  the factors associated with the rapidities 
$v_j$ and $v_k$  cancel each other. 
On the other hand, 
if $\ell > m$,  we have 
${v_{Pm} -v_{P\ell}+ 2i }$ in the enumerator of LHS of eq. (\ref{A1}),    
and we have 
\be
{\frac {v_{Pm} -v_{P\ell}+ 2i } {v_{j} -v_{k}+ 2i } }=
{\frac {v_{k} -v_{j}+ 2i } {v_{j} -v_{k}+ 2i } }. 
\label{me}
\ee
We can express these results  by the following 
$$
\left( {\frac {v_{k} -v_{j}+ 2i } {v_{j} -v_{k}+ 2i } }\right)^{H(\ell-m)} . 
$$
Considering  all the pairs $j,k$ with $j<k$, 
we  establish the equality (\ref{A1}).       
\begin{flushright}
{\it Q.E.D.} 
\end{flushright}

  \begin{prop}
  The amplitude $A_M(P)$ defined by eq. (\ref{AM1}) for  $P \in S_M$ 
  can be expressed as follows.  
  \be
  A_M(P) =  \prod_{1 \le j < k \le M} \left( 
  {\frac {v_j-v_k - 2i} {v_j-v_k +2i} } \right)^{H(P^{-1}j - P^{-1}k)} 
  \label{Lemma}
  \ee
  \end{prop} 
  ({\it Proof})
  The amplitude $A_M(P)$ defined by eq. (\ref{AM1}) is written 
  in terms of rapidities as follows 
 \be 
A_M(P) = \epsilon(P) \prod_{1 \le j<k \le M} 
{\frac {v_{Pj}-v_{Pk} + 2i } {v_{j}-v_{k} + 2i } }
\ee 
In Appendix C, we show the following identity in Prop. C.1 
\be 
\epsilon_M(P) = \prod_{1\le j<k \le M} 
(-1)^{H(P^{-1}j-P^{-1}k)}  
\ee
Thus,   making use of Lem. III.1 and Prop. C.1, 
we obtain   
\bea
A_M(P) & = & \epsilon(P) \prod_{1 \le j<k \le M} 
 {\frac {v_{Pj}-v_{Pk} + 2i } {v_{j}-v_{k} + 2i } } \non \\
& = & \epsilon(P) \prod_{1 \le j<k \le M} \left( 
{\frac {v_{k}-v_{j} + 2i }  {v_{j}-v_{k} + 2i } }
\right)^{H(P^{-1}j-P^{-1}k)}   \non \\
& = &  \prod_{1 \le j<k \le M} \left( 
{\frac {v_{j}-v_{k} - 2i }  {v_{j}-v_{k} + 2i } }
\right)^{H(P^{-1}j-P^{-1}k)}  .  
\eea
  \begin{flushright}
  {\it Q.E.D.}
  \end{flushright}

\par 
  We give a remark. Using  Prop. III.1, we can explicitly 
  prove that the Bethe states (and also the formal Bethe states) 
   should vanish when 
  there are two momenta of the same value.  The proof is 
  given in Appendix D. 

\subsection{Proof of the limit}

  Let us take a  permutation $P$ on $R+K$ letters 
  ($P \in {\cal S}_{R+K}$). We consider the following set 
  \be
  P^{-1}\{1,2, \ldots, R\}  =  
  \{ P^{-1}j \quad | \quad {\rm for} \quad j=1, 2, \ldots, R \}.
  \ee  
  Let us denote the elements of the set 
  by  $a_1, a_2, \ldots, a_R$, where     
  $a_j$'s are set in increasing order:   $a_1< a_2 < \cdots < a_R$.  
  For the permutation $P$,  we introduce  permutation $P_R$ on $R$ letters by  
  \be
  P_R m = P a_m \quad {\rm for} \quad m=1, \ldots, R.
  \label{PR}
  \ee
  Then, we have the following. 
  \begin{lem} Let $P_R$ denote the permutation on $R$ letters defined by (\ref{PR})
  for a given permutation $P$ on $R+K$ letters.  
  For two integers $j_1$ and $j_2$ with $1 \le j_1, j_2 \le R$, 
  the inequality 
  $P^{-1}j_1 < P^{-1}j_2$ holds
   if and only if 
  ${P_R}^{-1}j_1 < {P_R}^{-1} j_2$. Equivalently, we have 
  \be
  H(P^{-1}j_1 - P^{-1}j_2 ) = H({P_R}^{-1}j_1 - { P_R}^{-1}j_2 )
 \quad   
  {\rm for } \quad 1 \le j_1, j_2 \le R. 
  \ee  
  \end{lem}
  ({\it proof}) Let us denote 
  $P^{-1}j_1$ and   $P^{-1}j_2$ by $a_{m_1}$ and $a_{m_2}$, 
  respectively.  Then, by definition, we have 
  $m_1 = {P_R}^{-1} j_1$ and $m_2 = {P_R}^{-1} j_2$.   
  Here we recall that $a_j$'s are set in increasing order. Thus, we see that  
  $a_{m_1} < a_{m_2}$ if and only if $m_1 < m_2$, which gives the proof.   

\par 
Similarly, let us  introduce a permutation on $K$ letters. We consider 
the following set 
 \be
  P^{-1}\{R+1, R+2, \ldots, R+K \}  =  
  \{ P^{-1}j \quad | \quad {\rm for} \quad j=R+1, R+2, \ldots, R+K \}.
  \ee  
  We denote by $b_1, b_2, \ldots, b_K$, the elements of the above set. 
  Here  we assume that  $b_j$'s are in increasing order: 
  $b_1< b_2 < \cdots < b_K$.  
  We define permutation $P_K$ on $K$ letters by the following 
  \be
  P_K m = P b_m \quad {\rm for} \quad m=1, 2, \ldots, K.
  \label{PK}
  \ee
  Then, we can show  the following. 
  \begin{lem} 
  Let $P_K$ denote the permutation on $K$ letters defined by (\ref{PK}) 
  for a given permutation $P$ on $R+K$ letters. 
  For two integers $j_1$ and $j_2$ with $R+1 \le j_1, j_2 \le R+K$, 
  the inequality 
  $P^{-1}j_1 < P^{-1}j_2$ holds
   if and only if 
  ${P_K}^{-1}(j_1-R) < {P_K}^{-1} (j_2-K)$. Equivalently, we have 
  \be
  H(P^{-1}j_1 - P^{-1}j_2 ) = H({P_K}^{-1}(j_1-R) - { P_K}^{-1}(j_2-R) )
 \quad   
  {\rm for } \quad R+1 \le j_1, j_2 \le R+K. 
  \ee  
  \end{lem}

  Making use of Lemmas III.1, III.2 and III.3,  
  we now show the following proposition 
  \begin{prop} 
  Let us consider  two positive integers $R$ and $K$ satisfying $0< K \le L-2R$. 
  Let $v_1, v_2, \ldots, v_{R}$ be the 
  rapidities  of a given regular eigenvector  $|R \ra$ 
  with $R$ down-spins, and $v_{R+1}(\Lambda), \ldots, v_{R+K}(\Lambda)$
  be additional  $K$ rapidities which are given by 
  $v_{R+j}(\Lambda) = \Lambda + \delta_j$ for $j=1,2,\ldots, K$.
  Here $\delta_j$'s are arbitrary constants. 
  For the Bethe ansatz wavefunction 
   $f_{R+K}$ with its amplitudes $A_{R+K}(P)$'s given by 
   (\ref{AM1}),  
   we have the following limit. 
  \bea  & &  \lim_{\Lambda \rightarrow \infty } 
  f_{R+K}(x_1, \ldots, x_{R+K}; k_1, \ldots, k_R, k_{R+1}(\Lambda), \ldots,
  k_{R+K}(\Lambda)) \non \\
  & & = C_K \, \sum_{1 \le j_1 < \cdots < j_R \le R+K} 
  f_R(x_{j_1}, \ldots, x_{j_R}, k_1, \ldots, k_R) 
  \label{prop1}
 \eea 
  Here $k_j$'s  are related to the rapidities $v_j$'s through the relation: 
 $\exp i k_j = (v_j + i)/(v_j -i)$,  and the constant $C_K$ is given by 
 \be 
 C_K = \sum_{P \in S_K} A_K(P) \left[\delta_1, \cdots, \delta_K \right]
 \ee
  \end{prop}
  ({\it Proof})
   We recall that the Bethe ansatz wavefunction $f_{R+K}$ is given by 
  \be
  f(x_1,\ldots, x_{R+K}) 
  = \sum_{P \in S_{R+K}} A_{R+K}(P) \exp(i \sum_{j=1}^{R+K}k_{Pj} x_j) 
  \label{R+K}
  \ee
  Let us take a permutation $P$ in $S_{R+K}$.  
  By Lemma III.2 we can show that 
  the amplitude $A_{R+K}(P)$ of the formal Bethe state 
  is given by 
  \be 
  A_{R+K}(P)\left[v_1, \ldots, v_R, v_{R+1}(\Lambda), \ldots, v_{R+K}(\Lambda) \right] = 
  \prod_{1 \le  j < \ell \le R+K}
  \left( 
  {\frac {v_j- v_{\ell} -2i}{v_j- v_{\ell} + 2i}}
  \right)^{H(P^{-1}j-P^{-1}\ell)}
  \ee
  The above product can be decomposed into the three parts in the following 
  \bea
\prod_{1 \le  j < \ell \le R+K}
  \left( 
  {\frac {v_j- v_{\ell} -2i}{v_j- v_{\ell} + 2i}}
  \right)^{H(P^{-1}j-P^{-1}\ell)}
  & = & 
  \prod_{1 \le  j < \ell \le R}
  \left( 
  {\frac {v_j- v_{\ell} -2i}{v_j- v_{\ell} + 2i}}
  \right)^{H(P^{-1}j-P^{-1}\ell)} \non \\
  %
  & & \times 
  \prod_{1 \le  j \le R} \quad \prod_{R+1 \le \ell \le R+K}
  \left( 
  {\frac {v_j- v_{\ell} -2i}{v_j- v_{\ell} + 2i}}
  \right)^{H(P^{-1}j-P^{-1}\ell)} \non \\
  %
  & & \times 
  \prod_{R+1 \le j < \ell \le R+K}
  \left( 
  {\frac {v_j- v_{\ell} -2i}{v_j- v_{\ell} + 2i}}
  \right)^{H(P^{-1}j-P^{-1}\ell)} \non \\
   \label{123}
  \eea 
  First, we consider the third part of  RHS of (\ref{123}).
  Making use of Lemma III.3, we have 
  \bea   
  \prod_{R+1 \le j < \ell \le R+K}
  \left( 
  {\frac {v_j- v_{\ell} -2i}{v_j- v_{\ell} + 2i}}
  \right)^{H(P^{-1}j-P^{-1}\ell)} 
  & = & 
   \prod_{1 \le j < \ell \le K}
  \left( 
  {\frac {v_{j+R}- v_{\ell+R} -2i}{v_{j+R}- v_{\ell+R} + 2i}}
  \right)^{H(P_K^{-1}j-P_K^{-1}\ell)}  \non \\
 & = & 
   \prod_{1 \le j < \ell \le K}
  \left( 
  {\frac {\delta_{j}- \delta_{\ell} -2i}{ \delta_{j}- \delta_{\ell} + 2i}}
  \right)^{H(P_K^{-1}j-P_K^{-1}\ell)} 
  \label{third}
  \eea
  We note that  RHS of (\ref{third}) is nothing but  
  $A_K(P_K) \left[ \delta_1, \cdots, \delta_K \right]$.  
  Second,  it is clear that 
  the second part of RHS of (\ref{123}) becomes 1 under the limit: 
   $\Lambda \rightarrow \infty$.  
 In fact, putting the additional rapidities into the second part of  RHS of 
  (\ref{123}),   we have   
  \bea 
  & & 
  \prod_{1 \le  j \le R} \quad \prod_{R+1 \le \ell \le R+K}
  \left( 
  {\frac {v_j- v_{\ell} -2i}{v_j- v_{\ell} + 2i}}
  \right)^{H(P^{-1}j-P^{-1}\ell)} \non \\
  & =  & 
  \prod_{1 \le  j \le R} \quad \prod_{R+1 \le \ell \le R+K}
  \left( 
  {\frac {v_j- \Lambda -\delta_{\ell} - 2i}
  {v_j- \Lambda - \delta_{\ell} +2i}}
  \right)^{H(P^{-1}j-P^{-1}\ell)} . 
  \label{second} 
  \eea 
  Third, we consider the first part of RHS of (\ref{123}). 
    We recall that 
   ${P_R}$ is defined for the given permutation  $P$ by the relation (\ref{PR}). 
  Then, from Lemma III.2, we have 
  \be 
  \prod_{1 \le  j < \ell \le R}
  \left( 
  {\frac {v_j- v_{\ell} -2i}{v_j- v_{\ell} + 2i}}
  \right)^{H(P^{-1}j-P^{-1}\ell)} 
  = \prod_{1 \le j < \ell \le R}
  \left( 
  {\frac {v_j- v_{\ell} -2i}{v_j- v_{\ell} + 2i}}
  \right)^{H({ P_R}^{-1}j-{ P_R}^{-1}\ell)} 
  \label{first}
  \ee
  We note again  that RHS of (\ref{first}) is equal to 
  $A_R({P_R}) \left[ v_1, \cdots, v_R \right]$. 
 Thus, we have 
 \bea
& & \lim_{\Lambda \rightarrow \infty} 
A_{P+K}(P)\left[v_1, \cdots, v_R, 
v_{R+1}(\Lambda), \cdots, v_{R+K}(\Lambda) \right] 
\non \\
& =& A_{R}(P_R)\left[v_1, \cdots, v_R \right] \,  \times \, 
A_{K}(P_K)\left[\delta_1, \cdots, \delta_K \right]  
\eea
\par \noindent 
  Let us now consider the exponential part of (\ref{R+K}).  We note the following 
  \bea
  \sum_{j=1}^{R+K} k_{Pj}x_j & = & \sum_{\ell=1}^{R+K} k_{\ell} x_{P^{-1}\ell} 
  \non \\ 
  & = & \sum_{\ell=1}^{R} k_{\ell}x_{P^{-1}\ell} + 
  \sum_{\ell=R+1}^{R+K} k_{\ell}x_{P^{-1}\ell} . 
  \eea
 Since $k_{R+1} \ldots k_{R+K}$ are approaching to $0$ (mod $2\pi$)
 in the limit of sending $\Lambda$
to infinity, we have  
  \be 
  \lim_{\Lambda \rightarrow \infty}
   \sum_{\ell=R+1}^{R+K} k_{\ell}(\Lambda) x_{P^{-1}\ell} = 0 
   \quad ({\rm mod} \quad 2\pi). 
  \ee 
 Making use of the relation ${P_R}m = P a_{m}$, we have 
  \be 
   \sum_{\ell=1}^{R} k_{\ell}x_{P^{-1}\ell}  
  = \sum_{m=1}^{R} k_{{P_R}m}x_{P^{-1}{ P_R}m}  
  = \sum_{m=1}^{R} k_{{P_R}m}x_{a_m}  
  \ee
  Thus, we have 
  \bea 
  & & \lim_{\Lambda \rightarrow \infty} A_{R+K}(P) 
  \exp(\sum_{j=1}^{R+K} k_{Pj} x_j) 
   \non \\
  & = & 
A_K(P_K)\left[\delta_1, \ldots, \delta_K  \right] 
\times A_R(P_R)\left[ v_1, \ldots, v_R \right] 
\exp(\sum_{m=1}^{R} k_{{P_R}m} x_{a_m}) \quad {\rm for } \quad P \in S_{R+K} 
  \eea 
   Finally, we give a remark. To pick up  
   a permutation $P$ on $R+K$ letters is equivalent 
   to do the procedures in the following: 
    we take a subset $\{ a_1,a_2, \ldots, a_R \}$  of the set of $R+K$ letters 
    1, 2, $\ldots$, $R+K$,  and specify  $P_R$ on $R$ letters 
    and $P_K$ on $K$ letters by (\ref{PR}) and (\ref{PK}), respectively. 
    Therefore, we have 
\be 
\sum_{P \in S_{R+K}}  = \sum_{ \{a_1, \ldots, a_R \} 
\subset \{1,2, \ldots, R+K \}  } 
\sum_{P_R \in S_R} \sum_{P_K \in S_K} 
\ee 
  Thus, we have the relation (\ref{prop1}),  where $a_m$'s correspond to  $j_m$'s.
  \begin{flushright}
  {\it Q.E.D.}
  \end{flushright}

\par 
 It is now clear that we obtain Theorem III.1  from Proposition III.2. 
 

  \setcounter{section}{3} 
  \setcounter{equation}{0} 
 \renewcommand{\theequation}{4.\arabic{equation}}
 \section{Formal Bethe state from the algebraic Bethe ansatz}

\subsection{The algebraic Bethe ansatz for the XXX model under the P.B.C.s }

The formal Bethe state has been formulated  in terms of  
the coordinate Bethe ansatz method in sec. 1. 
However, in this section, we show  
that it is also important in the context 
of the algebraic Bethe ansatz method. 
In fact, we show that the formal Bethe state $|| M \ra $ 
with $M$ rapidities corresponds to the state 
created by the $B$ operators with the same set of rapidities.  
In the derivation, we use the method of the generalized two-site model 
first discussed by Izergin and Korepin 
\cite{Izergin-Korepin,Izergin-Korepin-Reshetikhin}.

\par 
Let us formulate some notation for the algebraic Bethe ansatz 
of the XXX model under the periodic boundary conditions.  
We define the $L$ operator acting on the $n$th 
site of the one-dimensional lattice by 
\be 
L_n(\lam) = 
\left( 
\begin{array}{cc} 
\lam I_n + \eta \sigma_n^{z} & 2 \eta \sigma_n^{-} \\
2 \eta \sigma_n^{+} &  \lam I_n - \eta \sigma_n^{z}  
\end{array} 
\right)
\ee
Here we recall that the $\sigma_n^z$'s are the Pauli matrices acting on the $n$th site. 
The monodromy  matrix is defined by the product of $L$ operators  
\be 
T(\lam) = L_L(\lam-q_L) L_{L-1}(\lam-q_{L-1}) \cdots L_1(\lam-q_1) 
\ee
Here the free variables $q_k$'s are called inhomogeneous parameters. 
Let us  denote the operator-valued matrix elements of the monodromy matrix by  
\be 
T(\lam) = \left( 
\begin{array}{cc} 
A(\lam) & B(\lam) \\
C(\lam) & D(\lam) 
\end{array} \right)
\ee
The transfer matrix of the XXX model is given 
by the trace of the monodromy matrix  
\be 
\tau(\lam) = {\rm tr} \left(T(\lam)\right) = A(\lam) + D(\lam) 
\ee 
The  Hamiltonian of the XXX model is  derived from  
the logarithmic derivative of the homogeneous transfer matrix  
where all the inhomogeneous parameters are given by zero. We note that 
when $q_k = 0$ for $k=1, \ldots, L$, 
we call the transfer matrix homogeneous.  
Explicitly, we have 
\be
\tau(\eta)^{-1} \left. {\frac d {d \lam}}  \tau(\lam) 
 \right|_{\lam=\eta; \, q_1=\cdots = q_L=0 }  ={ \frac 1 {4 \eta}} 
\sum_{j=1}^{L} \left( {\vec \sigma}_j \cdot {\vec \sigma}_{j+1} + 1 \right)
\ee

\par 
Let us consider the Yang-Baxter equations. 
With the $R$ matrix 
\be 
{R}(\lam) = 
{\frac 1 {\lam}} 
\left(
\begin{array}{cccc}
\lam + 2 \eta & 0 & 0 & 0 \\ 
0 & 2 \eta  &  \lam & 0 \\ 
0 &  \lam & 2 \eta & 0 \\ 
0 & 0 & 0 & \lam + 2\eta \\ 
\end{array}
\right) \, , 
\ee
we can show the Yang-Baxter equations for the $L$ operators  
\be 
{R}(\lam - \mu) \left( L_n(\lam) \otimes L_n(\mu) \right) 
= \left( L_n(\mu) \otimes L_n(\lam) \right) {R}(\lam - \mu) 
\label{RLL}
\ee
Here the symbol $\otimes$ denotes the direct product of matrices.  
Applying eq. (\ref{RLL}) to  each site,  
we can derive 
the Yang-Baxter equations for the monodromy matrix 
\be 
{R}(\lam - \mu) \left( T(\lam) \otimes T(\mu) \right) 
= \left( T(\mu) \otimes T(\lam) \right) {R}(\lam - \mu) 
\label{RTT}
\ee
The Yang-Baxter equations for the monodromy matrices  
 give  the set of commutation relations among the 
operators $A(\lam), B(\lam), C(\lam), D(\lam)$. 
For instance, we have
\bea 
A(\lam_1) B(\lam_2) & = & f_{12} B(\lam_2) A(\lam_1) - g_{12} B(\lam_1) A(\lam_2) \label{AB} \\
D(\lam_1) B(\lam_2) & = & f_{21} B(\lam_2) C(\lam_1) - g_{21} B(\lam_1) D(\lam_2) \label{DB} 
\eea
Here the symbols $f_{12}$ and $g_{12}$ denote the following 
\be 
f_{12} = {\frac {\lam_1 - \lam_2 - 2 \eta} {\lam_1 -\lam_2}} \, , 
\quad g_{12} = - {\frac {2 \eta} {\lam_1 -\lam_2}} 
\ee

The  operators $A(\lam)$ and $D(\lam)$ act on the vacuum $|0 \ra$ as  
\be 
A(\lam) |0 \ra =  a(\lam) \,|0 \ra \, , 
\quad D(\lam) |0 \ra =  d(\lam) \, |0 \ra
\ee
where $a(\lam)$ and $d(\lam)$ are given by 
\be 
a(\lam)  =  \prod_{k=1}^{L} (\lam - q_k + \eta) \, , \quad 
d(\lam)  =  \prod_{k=1}^{L}  (\lam - q_k - \eta)   
\ee

\subsection{The Bethe ansatz wavefunction from the generalized two-site model}

Let us explicitly calculate the entries (or matrix elements) 
of the vector 
$B(\lam_1) \cdots B(\lam_M) | 0 \ra$. 
Let us consider two sub-lattices of the one-dimensional lattice 
with sites 1 to $L$:  one consisting of sites 1 to $n$ and the other 
 of $n+1$ to $L$. 
The monodromy matrix $T(\lam)$ for the total lattice is given by the  
product of  two monodromy matrices for the sub-lattices   
\be 
T(\lam) = T^{(2)}(\lam) T^{(1)}(\lam) \label{TT} 
\ee
where $T^{(1)}(\lam)$ and  $T^{(2)}(\lam)$  are given by  
\bea 
T^{(1)}(\lam) & = & L_{n}(\lam-q_n)L_{n-1}(\lam-q_{n-1}) \cdots L_1(\lam-q_1)  
\non \\
T^{(2)}(\lam) & = & L_{L}(\lam-q_L)L_{L-1}(\lam-q_{L-1})
 \cdots L_{n+1}(\lam-q_{n+1})  
\eea
Let us denote the elements of the monodromy matrices of the sub-chains 
by the following 
\be 
T^{(\al)}(\lam) = \left( 
\begin{array}{cc} 
A^{(\al)}(\lam) & B^{(\al)}(\lam) \non \\
C^{(\al)}(\lam) & D^{(\al)}(\lam)
\end{array} \right) \quad {\rm for} \quad \al = 1,2 . 
\ee
Then, the product is given by 
\be 
\left( 
\begin{array}{cc} 
A(\lam) & B(\lam) \non \\
C(\lam) & D(\lam)
\end{array} \right) 
= \left( 
\begin{array}{cc} 
A^{(2)}(\lam) & B^{(2)}(\lam) \non \\
C^{(2)}(\lam) & D^{(2)}(\lam)
\end{array} \right) 
 \left( 
\begin{array}{cc} 
A^{(1)}(\lam) & B^{(1)}(\lam) \non \\
C^{(1)}(\lam) & D^{(1)}(\lam)
\end{array} \right) 
\ee
It gives four relations. Among them, we consider the following 
\be 
B(\lam) = A^{(2)}(\lam) B^{(1)}(\lam) + B^{(2)}(\lam) D^{(1)}(\lam)   
\label{two-site}
\ee
By applying the formula (\ref{two-site}) extensively,  
we can show the following \cite{Izergin-Korepin,Korepin-book}
\be 
\prod_{j=1}^{M} B(\lam_j) | 0 \ra = 
\sum_{S_{1}, S_{2}}^{((2))}  
\left(  \prod_{ k \in S_{1} } B_{k}^{(1)} 
\prod_{m \in S_{2} } B_{m}^{(2)}  \right) |0 \ra 
\left( \prod_{k \in S_{1}}   \prod_{m \in S_{2}}  f_{km} \right)  
\left( \prod_{k \in S_{1}} a_{k}^{(2)} \right)
\left( \prod_{m \in S_{2}} d_{m}^{(1)} \right)
\label{two}
\ee
Here the symbol $\sum_{S_{1}, S_{2}}^{((2))}$  denotes 
the sum over all the combination of sets 
 $S_1$ and $S_2$  satisfying the condition: 
 $S_{1} \cup S_{2}=\{1, \ldots, M\}$. 
Here we have denoted $B(\lam_k)$ by $B_k$, briefly. 
The symbol $a_k^{(2)}$ has been  defined by the relation:
$A^{(2)}(\lam_k) |0 \ra = a_k^{(2)}$, and so on. 
We can prove the formula (\ref{two}) by induction on $M$.

\par 
Let us consider the case where  the total lattice of 1 to $L$ 
is divided into $r$ parts. Then, 
applying the formula (\ref{two}), we can show the following 
\cite{Izergin-Korepin,Korepin-book} 
\bea
\prod_{j=1}^{M} B(\lam_j) | 0 \ra & = & 
\sum_{S_1, \ldots, S_r}^{((r))} \left( 
\prod_{\alpha=1}^{r} \prod_{ m_{\alpha} \in S_{\alpha} } 
B_{m_{\alpha}}^{(\alpha)} \right) | 0 \ra \non \\
& & \times \prod_{ 1 \le \alpha < \beta \le r} 
\Bigg\{ \left( \prod_{k \in S_{\alpha}}  
 \prod_{m \in S_{\beta}} f_{km} \right)  
\left( \prod_{k \in S_{\alpha}} a_{k}^{(\beta)} \right)
\left( \prod_{m \in S_{\beta}} d_{m}^{(\alpha)} \right)
\Bigg\}
\label{multi}
\eea
Here the symbol
$\sum_{S_1, \ldots, S_r}^{((r))}$ denotes 
the sum over all the sets $S_1, \ldots, S_r$ which satisfy the condition: 
$S_1 \cup S_2 \cup \cdots \cup S_r = \{ 1, 2, \ldots, M \}$.   
The formula (\ref{multi}) can be proven by induction on $r$.

\par 
Let us consider the case of $r=L$ where 
each of the sub-lattices is consisting of only one site. 
Then,  we can show  $B_k^{(\al)}B_m^{(\al)}| 0 \ra = 0$, 
for any $k,m \in S_{\al}$. Thus, if any one of the sets 
$S_1, \ldots, S_L$ contains more than one integer, 
then the contribution for the sets 
vanishes in eq. (\ref{multi}).   
Therefore, we may consider only such partitions of 
$\{1, \ldots, M\}$ into $L$ sets: $S_1, \ldots, S_L$ where 
each of $S_j$'s contains at most  one integer. 
Let us take one such partition and assume 
that the symbols $S_{x_1}, \ldots, S_{x_M}$ 
denote the non-empty sets  
with $1 \le x_1 < x_2 < \cdots < x_M \le L$. 
Then, there exists an element of the permutation group ${\cal S}_M$ 
such that  $S_{x_j}=\{ Pj \}$ for $j =1, \ldots, M$.  Now we can derive 
the following expression from eq. (\ref{multi}). 
\bea
\prod_{j=1}^{M} B(\lam_j) | 0 \ra & = & (2\eta)^M 
\sum_{P \in {\cal S}_M} \sum_{ 1 \le x_1 <  \cdots < x_M \le L} 
\prod_{j=1}^{M} \sigma^{-}_{x_j} | 0 \ra \non \\ 
& & \times \left( \prod_{ 1 \le \alpha < \beta \le M} 
 f_{Pk \, Pm} \right)  
\prod_{\alpha=1}^{M} 
\left( 
\prod_{1 \le  k < j_{\alpha} } d_{P \alpha }^{(k)} 
\prod_{j_{\alpha} < k \le L} a_{P \alpha}^{(k)} \right)
\label{wavefunction}
\eea
Here we have used the following 
\be 
\prod_{j=1}^{M} B_{Pj}^{(x_j)} | 0 \ra = (2 \eta)^M  
\prod_{j=1}^{M} \sigma_{x_j}^{-} |0 \ra 
\ee
which is valid when each of the $L$ sub-lattices  consists of one site. 
Here we note that the product $\prod_{1 \le j < k \le M} f_{Pj \, Pk}$  is related to 
the amplitudes of the Bethe ansatz wavefunctions 
\be 
\prod_{1 \le j < k \le M} f_{Pj \, Pk}
 =  \left( \prod_{1 \le j < k \le M} f_{j k}  \right)
\prod_{1 \le j< k \le M}
 \left( 
{\frac {\lam_j-\lam_k + 2 \eta }  
       {\lam_j-\lam_k - 2 \eta  }}  
\right)^{H(P^{-1}j-  P^{-1}k)}
\label{connection}
\ee

\par 
For the homogeneous case where $q_k = 0  $ for $k=1, \ldots , L$, 
we have 
\be 
\prod_{\al=1}^{M} 
\left( 
\prod_{1 \le  k < j_{\alpha} } d_{P \alpha }^{(k)} 
\prod_{j_{\alpha} < k \le L} a_{P \alpha}^{(k)} \right) =
\prod_{\al=1}^{M} 
\left\{ 
{\frac {(\lam_{P\al} + \eta)^L} {\lam_{P\al} - \eta}} 
 \left( {\frac {\lam_{P\al} - \eta} 
               {\lam_{P\al} + \eta} } \right)^{x_{\al}} 
\right\} 
\ee
Putting $\eta=-i$, we have 
\be
\prod_{j=1}^{M} B(\lam_j) | 0 \ra  = F_1 (\{ \lam_j \}) \,
 \sum_{P \in {\cal S}_M} \sum_{ 1 \le x_1 <  \cdots < x_M \le L} 
\prod_{j=1}^{M} \sigma_{x_j}^{-} | 0 \ra 
\left( \prod_{ 1 \le \alpha < \beta \le M} 
 f_{Pk \, Pm} \right)  
\exp \left(i \sum_{j=1}^{M} k_{Pj} x_j \right)  
\ee
where the factor $F_1(\{ \lam_j \}) $ has been given by 
\be 
F_1 (\{ \lam_j \}) = (2 \eta)^M \, 
\prod_{j=1}^{M} {\frac {(\lam_j-i)^L} {\lam_j +i}} 
\label{result} 
\ee
Noting the relation (\ref{connection}) we obtain 
\bea
\prod_{j=1}^{M} B(\lam_j) | 0 \ra &  = & F_1(\{ \lam_j \})  F_2(\{ \lam_j \}) 
  \non \\
& & \times  \sum_{P \in {\cal S}_M} \sum_{ 1 \le x_1 <  \cdots < x_M \le L} 
\left( \prod_{j=1}^{M} \sigma_{x_j}^{-} | 0 \ra \right) \, 
A_M(P)[\lam_1, \cdots,\lam_M]  \, 
\exp \left(i \sum_{j=1}^{M} k_{Pj}x_j \right)  
\eea
where the factor $F_2$ has been given by  
$ F_2(\{ \lam_j \}) = \prod_{1 \le j < k \le M} f_{j k} $.  
In terms of the notation of the formal Bethe state $|| M \ra $, 
we have  
\be 
B(\lam_1) \cdots B(\lam_M) | 0 \ra = 
 ||M \ra \, \times F_1 (\{ \lam_j \}) F_2(\{ \lam_j \})
\ee
Thus, we have shown that the formal Bethe state $|| M \ra $ is nothing but 
the vector $B(\lam_1) \cdots B(\lam_M) | 0 \ra$ where all the rapidities 
are given by free parameters.   Here we remark that 
the derivation of (\ref{wavefunction}) for the cases of $M=1$ and $M=2$ 
has already been given explicitly such as in Refs. \cite{Essler1,PR}.

\par 
We give some remarks.  The expression (\ref{wavefunction}) also holds for 
the XXZ model, where $f_{jk}$, $a(\lam)$ and $d(\lam)$  are replaced  by 
\bea 
f_{jk} & = & {\frac {\sinh(\lam_j -\lam_k - 2\eta)} {\sinh(\lam_j -\lam_k)} } \, , 
\quad g_{jk} = - {\frac {\sinh(2\eta)} {\sinh(\lam_j -\lam_k)} } 
 \non \\
 a(\lam) & = & \prod_{k=1}^{L} \sinh(\lam-q_k + \eta)  \, , 
\qquad d(\lam) = \prod_{k=1}^{L} \sinh(\lam-q_k - \eta)  \, . 
\eea 
Corresponding to eq. (\ref{connection}), for the XXZ model we have  
\be 
\prod_{1 \le j < k \le M} f_{Pj \, Pk}
 =  \left( \prod_{1 \le j < k \le M} f_{j k}  \right)
\prod_{1 \le j< k \le M}
 \left( 
{\frac {f_{kj} }  
       {f_{jk} }}  
\right)^{H(P^{-1}j-  P^{-1}k)}
\label{XXZamp}
\ee
A similar relation with (\ref{wavefunction}) has also been derived 
for the algebraic Bethe ansatz 
of the elliptic quantum group \cite{Felder}, where 
$f_{jk}$ are expressed in terms of the elliptic theta functions.



  \setcounter{equation}{0} 
 \renewcommand{\theequation}{5.\arabic{equation}}
 \section{The spectral flow under the twisted B.C.s }

\subsection{Formal Bethe states under P.B.C.s}

Let us discuss the formal Bethe state  
from the viewpoint of the ground-state solution of the XXZ model
given by Yang-Yang \cite{Yang-Yang}.  
We consider  the case of the anti-ferromagnetic Heisenberg model. 
In this section we assume that $L$ is even. 
We note that the relation between momentum and rapidity is 
slightly different from that of sec.1 and sec. 2 
due to the gauge transformation \cite{Yang-Yang}.   

\par 
The Hamiltonian of the XXZ model is given by 
\be 
{\cal H} = -J \sum_{\ell=1}^{L} \left(
 {S}_{\ell}^{x}  { S}_{\ell +1}^{x} + 
 { S }_{\ell}^{y}  { S}_{\ell +1}^{y} +
 \Delta { S }_{\ell}^{z}  { S}_{\ell +1}^{z}
\right) . \label{XXZ}
\ee 
where  $\Delta$ is called  the anisotropy parameter. 
Let us discuss the anti-ferromagnetic XXX model. 
Hereafter we assume $\Delta=-1$.   
Under the periodic boundary condition: 
${\vec S}_{L+1}={\vec S}_1$, we have the
Bethe ansatz equations 
\bea 
\exp(i L k_j ) & = &
(-1)^{M-1} \prod_{\ell=1, \ell \ne j}^{M} 
{\frac 
 {\exp[i(k_{j}+k_{\ell})] +1 -2 (-1) \exp(ik_{j}) }
 {\exp[i(k_j+k_{\ell})] +1 - 2 (-1) \exp(ik_{\ell})}  } , \non \\
& & \qquad  
{\rm for} \quad j=1,\ldots, M.
\label{PBA}
\eea 
We consider the case where  all the momenta $k_j$'s are real. 
Taking the logarithm of the Bethe ansatz  eqs. (\ref{PBA}), we have 
\be 
Lp_j = 2 \pi I_j - \sum_{\ell=1, \ell \ne j}^{M} 
\Theta(p_j, p_{\ell})  , \qquad 
{\rm for} \quad j=1, \ldots, M. 
\label{BAEreal}
\ee
Here $I_j= (M-1)/2$ (mod 1). The function 
$\Theta(p,q)$ has been given  by \cite{Yang-Yang}
\be 
\Theta(p,q)= 2 \tan^{-1} \left( 
{\frac {(-1) \sin((p-q)/2)}
{\cos((p+q)/2)- (-1) \cos((p-q)/2)}} 
\right)  
\ee
In Ref. \cite{Yang-Yang}, some important 
analytic properties of 
the function $\Theta(p,q)$ have been given. In particular, 
we notice the following:  
When    $-\pi < p < \pi$ and  $-\pi < q  < \pi$,  we have 
\bea 
\lim_{\epsilon \rightarrow 0, 0 < \epsilon  \ll 1  
} \Theta(\pi-\epsilon, q) & = & - \pi,  \qquad 
\lim_{\epsilon \rightarrow 0, 0 < \epsilon  \ll 1
} \Theta(-\pi+ \epsilon, q) =  \pi  ,  
\label{ppi} \\
\Theta(p,q) & = & -\Theta(q,p) .
\eea
Here we note that $\epsilon$ should be positive in the limits. 
In Ref. \cite{Yang-Yang}, the function is mainly discussed 
within the range $(-\pi, \pi)$ for $p$ and $q$.

\par 
Let us first show that 
the limit: $p \rightarrow -\pi$ under $p>-\pi$ corresponds 
to the limit of infinite rapidity: $\Lambda \rightarrow \infty$ 
discussed in the previous sections.  
Let us introduce the transformation between 
momentum $k$ and rapidity $v$ for the anti-ferromagnetic case 
($\Delta =-1$) 
\be 
\exp(ik) = (-1) \left( {\frac {v + i} {v -i}} \right) \quad . 
\label{kv}
\ee
Here the extra $(-1)$ factor corresponds to the gauge transformation, 
which is consistent with the periodic boundary condition 
if $L$ is even. 
By taking the logarithm of the relation (\ref{kv}) 
and choosing the branch of the logarithmic function such that 
is consistent with the function $\Theta(p,q)$, we have 
\be 
k = - 2 \tan^{-1} v . \label{kv-log}
\ee
It is clear that 
momentum $k$ is in the range $(-\pi, \pi)$ if  $v$ is finite. 
If we take the limit $v \rightarrow \infty$,  then 
the momentum $k$ approaches to $-\pi$ from the above 
($k > -\pi$).   
Thus, we can extend the range of real momenta  
as follows
\be 
-\pi \le k < \pi.  \label{newrange}
\ee

\par 
Let us call momentum $k_j$ {\it regular} 
if it satisfies the condition: $-\pi < k_j < \pi$. 
We  now introduce a symbol $T_1$ for $(L-M-1)/2$.  
From the viewpoint of the string-hypothesis, 
it is shown that the integer (or half-odd integer)
 $I_j$ for a regular real momentum $k_j$ 
satisfies the condition:
$|I_j| < T_1$ \cite{Bethe,TF,Takahashi,Takahashi-review}. 

\par 
We now consider such a  solution to the Bethe ansatz 
equations that has the momentum $k_0=-\pi$. 
Suppose that the set of $R$ regular momenta $k_1, \ldots, k_R$
and one non-regular momentum $k_0=-\pi$ gives a solution to 
the Bethe ansatz equations (\ref{BAEreal}) with $M=R+1$ down-spins.  
We have the following equations. 
\bea
L k_j & = & 2 \pi I_j - \sum_{\ell=0, \ell \ne j}^{R} 
\Theta(k_j, k_{\ell})  , \qquad 
{\rm for} \, \,  j=1, \ldots, R,  
\label{reg} \\ 
L k_0 &  = & 2 \pi I_0 - \sum_{\ell=1}^{R} 
\Theta(k_0, k_{\ell})  . 
\label{irr}
\eea
Here we have defined the equation for the momentum 
$k_0=-\pi$ by taking the limit 
of  $k$ to $-\pi$ under the condition $k > -\pi$. 
We also  assume that $I_j = R/2$ (mod 1) and 
$|I_j| \le T_1=(L-R)/2 -1$ for $j =1, \ldots, R$.  
($M-1=R$.)

\par 
Let us construct
 solutions to  eqs. (\ref{reg}) and (\ref{irr}). 
Making use of 
the limit (\ref{ppi}), 
we can easily show that $k_0=-\pi$ 
gives a solution to the equation (\ref{irr}) with 
$I_0 = - T_1 -1$, 
where $T_1 = (L-R)/2 -1$.  Recall that for $k_0=-\pi$, we assume 
the limit: $k_0 \rightarrow - \pi$ under the condition: $k_0 > -\pi$. 
From 
the property (\ref{ppi}),  we can show that the eq. (\ref{reg}) 
is equivalent to 
\be 
Lk_j  =  2 \pi (I_j +{ \frac 1 2})
- \sum_{\ell=1, \ell \ne j}^{R} 
\Theta(k_j, k_{\ell})  , \qquad 
{\rm for} \, \,  j=1, \ldots, R .   
\label{reg2} 
\ee
We note that the set of equations (\ref{reg2}) 
is equivalent to the standard 
Bethe ansatz equations (\ref{BAEreal}) 
for $R$ down-spins, where 
their numbers $(I_j+1/2)$'s satisfy the parity 
with $R$ down-spins: $I_j+1/2 = (R-1)/2$  (mod 1).  

\par 
Let us summarize the construction of solutions to 
(\ref{reg}) and (\ref{irr}). 
We first consider   the standard Bethe ansatz equations 
(\ref{BAEreal}) (or (\ref{reg2}))  with $R$ down-spins 
where their quantum numbers ${\hat I}_j$'s are given by 
 ${\hat I}_j=I_j +1/2$ for $j=1,\ldots, R$. We denote 
the solutions 
by ${\hat k}_1, \dots, {\hat k}_{R}$ 
for 
${\hat I}_1, \ldots, {\hat I}_R$, respectively. 
Then, the solutions to eqs. (\ref{reg}) and (\ref{irr})
are given by 
\bea 
{k}_j &  = & {\hat k}_j, \quad  
{\rm for } \,\, j =1, \ldots, R, \non \\ 
{k}_0 &  = & -\pi ,  
\eea
where ${I}_0= -(L-R)/2$.

\par 
It is interesting to note  that 
the set of momenta $k_0, \ldots, k_R$, 
constructed in the above 
corresponds to such a solution of the Bethe ansatz 
equations that gives  the non-regular eigenstate $|R,1 \ra$. 
In fact, we obtain the non-regular eigenstate $|R,1 \ra$,  
by taking the limit of the formal Bethe state $||R, 1; \Lambda \ra$ 
with the momenta $k_0, \ldots, k_R$, 
where $k_0$ corresponds to the infinite rapidity.

\subsection{Bethe states under the twisted B.C.s with a small twist}

Let us now consider the twisted boundary conditions
\be
S_{L+1}^{\pm} = S_{1}^{\pm} \exp(\pm i \Phi), 
\quad 
S_{L+1}^{z}=S_{1}^{z} . 
\label{twistedBC}
\ee
Here we call the variable $\Phi$ the twisting parameter. 
The Bethe ansatz equations under the twisted boundary conditions are given by 
\cite{Byers,deVega,Alcaraz,SS,Fowler,Kusakabe,Fukui} 
\bea 
\exp(i L p_j ) & = &
(-1)^{M-1} \exp(i \Phi) 
\prod_{\ell=1, \ell \ne j}^{M} 
{\frac 
 {\exp[i(p_{j}+p_{\ell})] +1 -2 (-1) \exp(ip_{j}) }
 {\exp[i(p_j+p_{\ell})] +1 - 2 (-1) \exp(ip_{\ell})}  } , \non \\
& & \qquad  
{\rm for} \quad j=1,\ldots, M.
\label{twistBAE}
\eea 
Let us assume that all the momenta $p_j$'s are real. 
Taking the logarithms of eqs. (\ref{twistBAE}), we have 
\be 
Lp_j = 2 \pi I_j + \Phi - \sum_{\ell=1, \ell \ne j}^{M} 
\Theta(p_j, p_{\ell}) , \quad  
{\rm for} \quad j=1,\ldots, M.
\label{twistBAEreal}
\ee
where $I_j= (M-1)/2$ (mod 1).

\par 
Let us discuss the adiabatic behaviors of 
 solutions of the Bethe ansatz equations 
(\ref{twistBAEreal}) 
under the twisted boundary conditions,  
where  $\Phi$ is a very small positive number, and is  
increased adiabatically. 
Here we assume that  $R+1$ momenta: $p_0, p_1, \ldots, p_R$ 
 are  solutions of  the Bethe ansatz equations  
where one momentum $p_0$ is close to 
$-\pi$ ($p_0 \sim -\pi$). We also assume that $R < L/2$. 
We consider the following equations. 
\bea
Lp_j & = & 2 \pi I_j + \Phi - \sum_{\ell=0, \ell \ne j}^{R} 
\Theta(p_j, p_{\ell})  , \qquad 
{\rm for} \, \,  j=1, \ldots, R,  
\label{reg-tw} \\ 
Lp_0 &  = & 2 \pi I_0 + \Phi - \sum_{\ell=1}^{R} 
\Theta(p_0, p_{\ell})  . 
\label{irr-tw}
\eea
Here,  
 $I_j = R/2$ (mod 1) and 
$|I_j| \le T_1=(L-R)/2 -1$ for $j =1, \ldots, R$.  
(Note $M-1=R$.) Now, 
let us introduce  a small positive number $\epsilon$    
and express $p_0$  as  $p_0 = -\pi +\epsilon$. 
Then, we can expand the function $\Theta(p,q)$ in the following 
\bea
\Theta(-\pi+\epsilon, q) &  = &  
\pi - 2 \epsilon + \epsilon^2 \tan {\frac q 2} + O(\epsilon^3), \non \\ 
\Theta(\pi-\epsilon, q) & = & 
- \pi + 2 \epsilon + \epsilon^2 \tan {\frac q 2} + O(\epsilon^3). 
\label{expansion}
\eea
Substituting the expansion into eq. (\ref{irr-tw}) we have 
\be 
2\pi I_0 +\Phi = -\pi(L-R) + \epsilon (L-2R) + O(\epsilon^2). 
\ee
Let us consider the adiabatic change of the twisting parameter 
$\Phi$. 
When we  change the parameter $\Phi$ infinitesimally, 
then the quantum number $I_0$ does not change at all. 
Thus, we obtain the following solutions 
\bea
I_0 & = & -  {L-R}/ 2 = -T_1 -1, \non \\
\epsilon & = &  {\Phi}/( {L-2R}) + O(\epsilon^2).  
\eea
The solution: $p_0= -\pi + \Phi/(L-2R) + O(\epsilon^2)$ 
can be considered as a regular solution, since it satisfies 
the condition $-\pi < p_0 < -\pi$. 
Furthermore, it is clear that 
$p_0$ approaches to $k_0$ under the limit 
$\Phi \rightarrow 0$ with $\Phi > 0$. 
Therefore, we conclude  that 
the formal solution 
$k_0=-\pi$ under the periodic boundary conditions 
corresponds to the regular solution 
$p_0= -\pi + \Phi/(L-2R) +O(\epsilon^2)$ 
under the twisted boundary conditions. 

\par 
We make a remark on the number of solutions to eq. (\ref{twistBAEreal}). 
Under the periodic boundary conditions, 
there are $2T_1+1=L-M$ regular solutions. 
Under the twisted boundary conditions,
 however, we have one more regular solution and 
$2T_1 +2= L-M+1$ regular solutions in total.

\subsection{ Analytic continuation of $\Theta(p,q)$ and $2\pi$-shift of 
momentum} 

In the previous subsection we have considered the case where the 
twisting  parameter $\Phi$ is very small. 
Hereafter, we discuss how the ground-state solution  
changes with respect to the parameter $\Phi$, and then we shall 
show that the spectral flow should have 
the period of $4\pi$, at the end of sec. 5. 

\par 
Let us consider an  extension of 
the function $\Theta(p,q)$ of the variable $p$ defined on  
the range $(-\pi,\pi)$ into a continuous function defined over 
$(-\infty , \infty)$.  Taking into account the analytic property (\ref{ppi}), 
we make the extension with respect to $p$ as follows 
\be
{\Xi}(p,q) = 
   \displaystyle  
  \left\{ 
  \begin{array}{c}
 - (2n+1)\pi,   \qquad  {\rm if } \quad  
  p=  (2n + 1) \pi,  \quad {\rm for} \, {\rm an} \, {\rm integer} \quad n, \\
      \displaystyle 
     \Theta(p- 2 \pi \left[
 {\frac {p + \pi} {2 \pi}} \right]  ,q) 
  - 2 \pi \left[
 {\frac {p + \pi} {2 \pi}} \right], 
 \quad {\rm otherwise}.
  \end{array}
  \right.
\label{ext}
\ee
Here the symbol $[x]$ denotes the Gauss'  symbol. 
We recall that $p$ is an arbitrary real number in eq. (\ref{ext}). 
We also extend the function $\Theta(p,q)$ 
with respect to $q$ by assuming the relation 
$\theta(p,q)=-\theta(q,p)$. 

\par 
In terms of the extended function, 
the Bethe ansatz equations are given by 
\be 
Lp_j = 2 \pi I_j + \Phi - \sum_{\ell=1, \ell \ne j}^{M} 
\Xi(p_j, p_{\ell}) , 
\quad  {\rm for} \quad j=1,\ldots, M, 
\label{xiBAEreal}
\ee
where $I_j= (M-1)/2$ (mod 1).  We recall that 
the range of $p_j$'s are given by the extended zone $(-\infty, \infty)$
in eqs. (\ref{xiBAEreal}). 
We note that for regular solutions, 
 eqs. (\ref{xiBAEreal}) 
are equivalent to the standard Bethe ansatz equations (\ref{twistBAEreal}). 
We recall that when $p_j$'s are regular, 
then   they satisfy $-\pi < p_j < \pi$.

\par 
Let us discuss the number of all the possible solutions  
to eqs. (\ref{xiBAEreal}). 
We assume that $p_1, \ldots, p_M$ are solutions 
to (\ref{xiBAEreal}) with the quantum numbers
$I_1, \ldots, I_M$, respectively. 
Let us  take a suffix $j_1$ from $1, \ldots,M$.  
We consider  momenta ${\hat p}_j$ given by 
\be  
{\hat p}_{j_1} = p_{j_1} +2\pi \quad {\rm and} \quad  
{\hat p}_j=p_j  \quad {\rm if} \quad j \ne j_1 \, (1 \le j \le M)  .  
\ee
Putting ${\hat p}_j$ 's into the eq. (\ref{xiBAEreal}), 
  we see that they give solutions to eq. 
(\ref{xiBAEreal}) with the quantum numbers 
${\hat I}_j$'s where they are given by 
\be 
{\hat I}_{j_1} = I_{j_1} + L-M+1 , \quad {\rm and } \quad   
{\hat I}_j = I_j + 1 \quad {\rm for} \quad j \ne j_1  \, (1 \le j \le M) . 
\label{shift}
\ee
Thus, we may regard the number  $L-M+1$  as the period of the quantum numbers 
$I_j$'s. Therefore,  we may consider only $L-M+1$ different values for 
$I_j$'s such as  $-T_1-1, -T_1, \cdots, T_1$. This choice is consistent with 
the range of momentum $-\pi \le k < \pi$. ($I_0=-T_1 -1$ corresponds to 
$k_0=-\pi$ or $v = \infty$ when $\Phi=0$.) 
Thus, if there is a one-to-one correspondence  between $k_j$'s and 
$I_j$'s, then the number of solutions 
of the Bethe ansatz equations is given by 
 $L-M+1$. We note $L-M+1= (2T_1 +1) +1$,  where $2 T_1 +1$ is 
the number of regular solutions given by the string hypothesis. 

\par 
Let us make a remark on  the change in the total momentum $P_{tot}$
induced by the shift: $p_{j_1} \rightarrow p_{j_1}+2\pi$.  
In fact, it 
is consistent with the shift of the quantum numbers ${\hat I}_j$'s. 
We denote by ${\hat P}_{tot}$ the total momentum for the 
momenta ${\hat p}_j$'s and ${\hat I}_j$'s. Then, we have 
\bea
{\hat P}_{tot} & = &
\sum_j {\hat p}_j = 2 \pi + \sum_j p_j , \non \\
 {\frac {2 \pi} L}\sum_j {\hat I}_j  & =&  
{\frac {2 \pi} L}\sum_{j \ne j_1} ({I}_j -1) + {\frac {I_{j_1} +L-M +1} L} 
=  2 \pi + {\frac {2 \pi} L} \sum_j {I}_j . \non \\ 
\eea

\subsection{Period $4 \pi$ of spectral flow under the twisted B.C.s}

Let us discuss how 
the ground-state solution changes 
when we increase the twisting parameter $\Phi$ 
adiabatically from 0 to $4 \pi$. In fact, it is known that 
 the spectral flow of the ground state energy has the period of 
$4 \pi$ with respect to the twisting parameter 
\cite{Alcaraz,SS,Fowler,Kusakabe,Fukui}.
In the following paragraphs, 
we shall show that the shift (\ref{shift})
on the $I_j$'s  is consistent with the $4 \pi$ period 
of the spectral 
flow of the ground-state energy.  

\par 
We consider the half-filling case where 
$M=L/2$. Here, we have $T_1=(L/2-1)/2$.  
Let symbols $p_j(\Phi)$ for $j=1,\ldots, M$ 
denote the momenta satisfying the Bethe ansatz equations
(\ref{xiBAEreal}) under the twisted boundary condition. 
We assume that 
when $\Phi =0$, the momenta are given by those of the 
ground state, where 
$-\pi < p_j(0)< \pi$ 
and $I_j=-T_1 +(j-1)$ for $j=1, \ldots, L/2$. 
We may assume that 
$p_1(0) < \cdots < p_{L/2}(0)$. 

\par 
Let us recall the adiabatic hypothesis that 
when we change the parameter $\Phi$ infinitesimally, 
then the quantum number $I_j$ for momentum $p_j(\Phi)$ does not 
change at all, while the momentum $p_j(\Phi)$ changes 
 proportionally to the infinitesimal change of $\Phi$. 
We note that 
under the adiabatic hypothesis we have 
$ k_1(\Phi) < \cdots < k_{L/2}(\Phi)$. 
Now we consider the case when $\Phi$ is very close to  $2\pi$. 
We set $\Phi= 2\pi -\delta$, where $\delta$ is a small positive number. 
Making use the expansion (\ref{expansion}), 
we can solve the Bethe ansatz equations (\ref{xiBAEreal}), and 
we obtain the following 
\be 
p_{L/2}(2\pi -\delta) = \pi - \delta/2 + O(\delta^2) . 
\ee
Thus, when $\Phi=2\pi$, we have 
momenta $p_1(2\pi), \ldots, p_{L/2}(2\pi)$, where  
their quantum numbers are increased by 1 and they are given by   
\be I_j=-T_1 +(j-1)+1 \quad {\rm  for} \quad  j=1, \ldots, L/2 . 
\ee
Here we note that $p_{L/2}(2\pi)= \pi$ and $I_{L/2}=T_1+1$. 
Let us now apply the shift (\ref{shift}) to the system of the solutions. 
We note that $p_{L/2}(2\pi) - 2 \pi = - \pi$ and 
$T_1+1 - (L-M+1)= - T_1-1$. 
Thus, they are equivalent to the following set 
of solutions to (\ref{xiBAEreal})
$$
p_{L/2}(2\pi)-2\pi=-\pi, \,\, p_1(2\pi), \ldots, p_{(L-2)/2}(2\pi)
$$
 with their quantum numbers given by 
$$
I_{L/2}=-T_1 -1, \,\, I_{1}=-T_1, \ldots, T_1 -1, 
$$
respectively.  
We further increase the twist parameter $\Phi$ until it becomes 
$4\pi$.  Then, we see that 
the set of momenta 
$$p_{L/2}(4\pi)-2\pi, \, \, p_1(4\pi), \ldots, p_{L/2-1}(4\pi)$$
 satisfy the
Bethe ansatz equations (\ref{xiBAEreal}) 
with 
$$I_{L/2}=-T_1, \, \, I_1=-T_1 +1, 
\cdots, I_{(L-2)/2}=T_1. 
$$ 
Thus, we obtain 
$$
p_{L/2}(4\pi)-2\pi=p_1(0), 
p_1(4\pi)=p_2(0), \ldots, p_{(L-2)/2}(4\pi)=p_{(L-2)/2}.
$$ 
The set of momenta for $\Phi=4\pi$ is equivalent to that of 
$\Phi=0$. Therefore we conclude that the  solutions to the 
Bethe ansatz equations which corresponds to the ground-state 
at $\Phi=0$ has the period of $4\pi$.


  \setcounter{section}{5} 
  \setcounter{equation}{0} 
 \renewcommand{\theequation}{6.\arabic{equation}}
 \section{Discussions}
 
In this paper, we have explicitly shown  that 
 any non-regular eigenvector is derived from the Bethe ansatz 
 wavefunction with infinite rapidities, 
for the one-dimensional XXX model under the periodic boundary conditions. 
The formula (\ref{amp}) for the amplitudes 
of the Bethe ansatz wavefunction has played a central role  in the proof. 

\par 
Let us discuss the string hypothesis, explicitly.  
It is based on the  assumption  
that the Bethe ansatz equations (\ref{BAEr}) have  
complex solutions given in the following 
\be 
v_{\alpha}^{n,j} = v_{\alpha}^n + i (n+1 -2j) 
+ \epsilon_{\alpha}^{n,j}  \quad 
{\rm for} \quad j=1, \ldots, n. 
\label{string}
\ee 
Here, it is also assumed that the absolute values of the 
correction terms $|\epsilon_{\alpha}^{n,j}|$ should be  very small.  
The set of complex rapidities  $v_{\alpha}^{n,j}$ for $j=1,\ldots, n$ 
is called an $n$-string solution \cite{Bethe,Takahashi,Kirillov} . 
The value $v_{\alpha}^n$ is called the  {\it center} of the string solution. 
The number $n$ is called the length of the string solution. 

\par 
Let us discuss the formula (\ref{amp}) of the amplitudes 
from the viewpoint of the string hypothesis.   
For  any given $n$-string solution, 
we set the $n$ rapidities in the string 
in such an order that $v_{\al}^j - v_{\al}^k \approx 2i (k-j)$ for any $j< k$ with 
$1 \le j, k \le n$.  
Then, the value of the amplitude $A_M(P)$ given by eq. (\ref{amp}) 
becomes stabilized and well-defined, since we can avoid the appearance of 
any very small factor of  $O(\epsilon)$  in the denominator of eq. (\ref{amp}).

\par 
Let us discuss the number of solutions to the Bethe ansatz equations 
 for the strings of length $n$ \cite{Bethe,Takahashi} .
Let us define number $T_n$ by 
\be 
T_n= {\frac 1 2} \left(N-1-\sum_{m=1}^{\infty} t_{nm} M_{m}
\right)
\label{Takahashi-number}
\ee
Here $M_m$ denotes the number of string solutions of length $m$ and $t_{nm}$ 
is given by 
$$ 
t_{nm}= 2 {\rm min}(n,m) - \delta_{nm} \, . 
$$
Under the periodic boundary conditions 
($\Phi=0$), it is discussed  that 
the number of string solutions of length $n$ is given by 
 $2T_n+1$ \cite{Takahashi,Kirillov} . We can show  that 
if there are $2T_n+2$ different string solutions of length $n$, 
then all the  solutions 
to the Bethe ansatz 
equations correspond to a complete set of the eigenvectors of 
the XXX Hamiltonian under the twisted boundary conditions. 
The result of the present paper 
suggests that the $K$ infinite 
rapidities of a non-regular eigenvector $|R,K \ra$ might correspond to 
a $K$-string solution under the twisted boundary conditions. 
Thus, we have a conjecture that 
any non-regular eigenvector under the P.B.C.s of the form $|R,K \ra$ 
should correspond to a regular eigenvector with  a $K$-string solution 
under the twisted B.C.s.  
It seems that the conjecture  should be consistent with the result of Ref. 
\cite{Tarasov-Varchenko}. However, a detailed numerical research on 
$K$-strings with large $K$'s should be performed such as studied  
in Ref. \cite{Shastry}.

\par 
Finally, we give a remark on a possible application of the result of 
the present paper 
to the XXZ and XYZ models. 
Recently, it has been shown that under the periodic boundary conditions, 
the one-dimensional XXZ Hamiltonian  at the $q$ root of unity conditions 
 has the $sl_2$ loop algebra symmetry \cite{sl2loop,XXZincomplete,complete}. 
In fact, we can discuss the spectral 
degeneracy of the XXZ model at the root of unity conditions 
in terms of  the algebraic Bethe ansatz method by applying 
some of the techniques developed in the paper \cite{TD}: 
combining the expression (\ref{wavefunction}) with the formula (\ref{XXZamp}) of the amplitudes  
$A_M(P)$'s, we can construct singular solutions related to the $sl_2$ loop algebra. 
Thus,  we can show  the validity of  the construction of 
the complete $N$-string solutions discussed in Ref. \cite{complete} 
in the level of eigenvectors. We can also prove it by showing 
that the limits of the  Bethe ansatz wavefunctions 
satisfy the sufficient conditions for  the eigenvectors of the XXZ model,  
which  are summarized in Appendix E.   
Surprisingly,  a similar method  can also be applied 
to the analysis of the spectral degeneracy of the XYZ model 
addressed in Ref. \cite{sl2loop}.     
The details will be discussed in subsequent papers.

{\vskip 1.2cm }
\par \noindent 
{\bf Acknowledgement}

The author would like to thank  
Profs. V.E. Korepin and B.M. McCoy 
for helpful discussions and valuable comments during the author's one-year stay 
 at Stony Brook in 1998-1999, where some part of the content of the paper
has been obtained.  
He would also like to thank Drs.  K. Kusakabe and R. Yue
for fruitful collaboration on Ref. 
\cite{boundary-states}  which gives a strong motivation of 
the present work. He is thankful to Prof. T. Nakanishi 
for useful comments on Ref. \cite{Tarasov-Varchenko} during the 
conference MATHPHYS ODYSSEY 2001. 
He is grateful to Dr. P.K. Ghosh 
for an interest in this work. 
This work is partially supported by the Grant-in-Aid for Encouragement 
of Young Scientists (No. 12740231).


\setcounter{section}{0}
\renewcommand{\thesection}{\Alph{section}}



\setcounter{equation}{0} 
 \renewcommand{\theequation}{A.\arabic{equation}}
\section{Appendix:  Formula for the action of spin-lowering operator } 

\par 
Let us introduce some  symbols. First, 
we shall abbreviate the symbol $\sum_{1 \le x_1 < \cdots < x_M \le L}$ 
 by $\sum_{x_1 < \cdots < x_M}^{\sim}$, in short.   
Second, for a non-negative integer $K$, we denote by 
the symbol $\sum_{\{j_1,j_2, \ldots, j_M \} 
\subset \{ 1, \ldots, M+K \} }$   
 the summation over all the subsets $\{j_1,j_2, \ldots, j_M \}$   
of $\{ 1, 2, \ldots, M+K \}$, 
where $j_k$'s are set in increasing 
order $j_1 < \cdots < j_M$.   Thus, 
the two symbols in the following express the same sum.  
\be 
\sum_{ \{ j_1, \ldots, j_M \} \subset \{ 1, \ldots, M+K \}} =  
\sum_{1 \le j_1 <  \cdots < j_M \le M+K}  \, . 
\ee 

\begin{prop}
Recall that $|M)$ denotes an arbitrary vector with $M$ down-spins 
defined by eq. (\ref{|M)}). We denote by  $|M,K)$  the vector obtained 
from  $|M)$ multiplied by the power of the spin-lowering operator 
\be
|M,K) =  {\frac 1 {K!}} \left( S_{tot}^{-} \right)^K |M)  
\ee
Then, we can show the following formula  
\be 
|M,K) = 
\sum_{x_1 < \cdots < x_{M+K}}^{\sim} \left( 
\sum_{ \{ j_1, \ldots, j_M \} \subset \{ 1, \ldots, M+K \}} 
g(x_{j_1}, \ldots, x_{j_M}) \right) 
\sigma_{x_1}^{-} \sigma_{x_2}^{-} \cdots \sigma_{x_{M+K}}^{-} |0 \ra   
\label{SSS} 
\ee
\end{prop}
({\it Proof})
We prove the formula (\ref{SSS}) by induction on $K$. 
\par \noindent 
(i) We show (\ref{SSS}) for the case of $K=1$. 
Applying $S_{tot}^{-}$ to $|M)$, we have 
\bea 
S_{tot}^{-}|M) & =& \sum_{y=1}^{L} \sigma_y^- 
\sum_{x_1 < \cdots < x_M} g(x_1, \ldots, x_M) 
\sigma_{x_1}^{-} \cdots \sigma_{x_M}^{-} |0 \ra  \non \\ 
& =& \sum_{x_1 < \cdots < x_M}^{\sim} \left( 
\sum_{y=1}^{y < x_1}
+ \sum_{y > x_1}^{ y< x_2} + \cdots + \sum_{y > x_M}^{L}  \right) 
g(x_1, \ldots, x_M) 
\sigma_y^- \sigma_{x_1}^- \cdots \sigma_{x_M}^- |0 \ra . \non \\
\label{all}
\eea 
We note  the following calculation. 
\bea
& & \sum_{x_1 < \cdots < x_M}^{\sim} \sum_{y > x_j}^{ y < x_{j+1}}
g(x_1, \ldots, x_M) 
\sigma_y^- \sigma_{x_1}^- \cdots \sigma_{x_M}^-  \non \\
& = & \sum_{x_1< \cdots < x_j < y < x_{j+1} < \cdots x_M}^{\sim}  
g(x_1, \ldots, x_M) 
\sigma_{x_1}^- \cdots \sigma_{x_j}^- \sigma_y^- \sigma_{x_{j+1}}^- \cdots 
\sigma_{x_M}^-  \non \\
& = & \sum_{x_1< \cdots < x_{M+1} }^{\sim} 
g(x_1, \ldots, x_j, \underbrace{x_{j+2}, \ldots, x_{M+1}}_{(j+1){\rm th}, \ldots, M{\rm th}}) 
\sigma_{x_1}^{-} \cdots \sigma_{x_{M+1}}^{-}  
\label{each}
\eea
In the last line, we have replaced the symbols 
$y$, $x_{j+1}$, \ldots, and $x_{M}$ by 
 $x_{j+1}$, $x_{j+2}$, \ldots, and $x_{M+1}$, respectively. 
Substituting (\ref{each}) into (\ref{all}),  we have 
\bea 
 S_{tot}^{-} |M) & = & 
\sum_{x_1 < \cdots < x_{M+1}}^{\sim} \big( 
g(x_2, \ldots, x_{M+1}) + g(x_1, x_3, \ldots, x_{M+1}) + \cdots  \non \\ 
& & + \cdots + g(x_1, x_2, \ldots, x_{M}) \big) 
  \sigma_{x_1}^- \cdots \sigma_{x_{M+1}}^{-} |0 \ra   \non \\
& = & 
\sum_{x_1 < \cdots < x_{M+1}}^{\sim}  
\left( 
\sum_{\{j_1, \ldots, j_M\} \subset \{1,2, \ldots,M+1\} }
g(x_{j_1}, \ldots, x_{j_M}) 
\right) 
 \sigma_{x_1}^- \cdots \sigma_{x_{M+1}}^{-} |0 \ra   . \non \\
\eea
Thus, we have the expression (\ref{SSS}) for the case of $K=1$. 
\par \noindent 
(ii) Let us assume the expression 
(\ref{SSS}) for the case of $K$.  Then, we show the case of 
$K+1$ in the following. 
\bea 
& & S_{tot}^{-} |M,K) \non \\
& = & \sum_{y=1}^{L} \sigma_{y}^{-} 
\left( \sum_{x_1 < \cdots < x_{M+K}} 
\sum_{ \{j_1, \ldots, j_M \} \subset \{ 1, \ldots, M+K \}}
g(x_{j_1}, \ldots , x_{j_M}) 
\sigma_{x_1}^- \cdots \sigma_{M+K}^{-} \right)  |0 \ra  \non \\
& =& \sum_{x_1 < \cdots < x_{M+K} }^{\sim} \left( 
\sum_{y=1}^{y < x_1}
+ \sum_{y > x_1}^{ y < x_2} + \cdots + \sum_{y > x_{M+K} }^{L} 
 \right)  \non \\
& & \times \sum_{ \{j_1 \cdots j_M \} \subset \{ 1, \ldots, M+K \} }
g(x_{j_1}, \ldots, x_{j_M}) 
\sigma_y^{-} \sigma_{x_1}^{-} \cdots \sigma_{x_{M+K}}^{-} |0 \ra . \label{all2}
\eea 
By a similar method for the case (i), we can show the following 
\bea
& & \sum_{x_1 < \cdots < x_{M+K} }^{\sim} 
\sum_{y > x_{\ell} }^{y < x_{\ell+1} }
\sum_{ \{j_1, \ldots, j_M \} \subset \{ 1, \ldots, M+K \}}
g(x_{j_1}, \ldots, x_{j_M}) \quad 
\sigma_y^- \sigma_{x_1}^- \cdots \sigma_{x_M}^-  \non \\
& = & \sum_{x_1< \cdots < x_{\ell} < y < x_{\ell+1} 
< \cdots <x_{M+K} }^{\sim} \quad  
\sum_{ \{j_1, \ldots, j_M \} \subset \{ 1, \ldots, M+K \}}
g(x_{j_1}, \ldots, x_{j_M}) 
\sigma_{x_1}^- \cdots \sigma_{x_{\ell} }^- \sigma_y^- 
\sigma_{x_{\ell+1}}^- \cdots \sigma_{x_M}^-  \non \\
& = & \sum_{x_1< \cdots < x_{M+K+1} }^{\sim} \quad 
\sum_{ \{j_1, \cdots, j_M \} \subset
 \{ 1, \ldots,\ell,\ell+2,\ldots, M+K+1 \}}
 g(x_{j_1}, \ldots, x_{j_{M}}) 
 \sigma_{x_1}^{-} \cdots \sigma_{x_{M+1}}^{-} \label{each2}
\eea
In the last line, we have replaced the symbol 
$y$ by $x_{\ell+1}$, and  $\ell+1, \ldots, M$ by 
$\ell+2, \ldots, M+1$, respectively. 
Substituting (\ref{each2}) into (\ref{all2}), we have 
\bea 
& & S_{tot}^- |M,K) \non \\
 & = &
\sum_{x_1 < \cdots < x_{M+K+1}}^{\sim} \left( 
\sum_{ \{j_1, \cdots, j_M \} \subset \{ 2,3, \ldots, M+K+1 \}}
+ \sum_{ \{j_1, \cdots, j_M \} \subset \{ 1,3, \ldots,  M+K+1 \}}
+ \cdots + 
\sum_{ \{j_1, \cdots, j_M \} \subset \{ 1,2, \ldots, M+K\}}
\right) \non \\ 
& & \times  g(x_{j_1}, \ldots, x_{j_M}) 
  \sigma_{x_1}^{-} \cdots \sigma_{x_{M+K+1}}^{-} |0 \ra   \non \\
& = & 
(K+1) \sum_{x_1 < \cdots < x_{M+K+1}}^{\sim}   
\left( 
\sum_{\{j_1, \ldots, j_M\} \subset \{1,2, \ldots,M+K+ 1\} }
g(x_{j_1}, \ldots, x_{j_M}) 
\right) 
 \sigma_{x_1}^{-} \cdots \sigma_{x_{M+1}}^{-} |0 \ra   
\eea
In the derivation of the last line, we note  
 that after selecting $M$ integers $j_1, j_2, \ldots, j_M $
from the set $\{1, 2, \ldots, M+K+1 \}$,  
there are $(K+1)$ ways for 
choosing one more element from the remaining $K+1$ integers. 
Thus, we have the factor $(K+1)$.
\begin{flushright}
Q.E.D. 
\end{flushright}


\setcounter{equation}{0} 
 \renewcommand{\theequation}{B.\arabic{equation}}
\section{Appendix:  Formal Bethe state with three infinite rapidities} 

\par 
Let us discuss the infinite limit of the formal Bethe state 
$||R,K; \Lambda \ra$ of the case $R=0$ and $K=3$.  
Here, $v_1$, $v_2$ and $v_3$ are additional rapidities given by 
 $v_1= \Lambda + \delta_1$, $v_2= \Lambda + \delta_2$ and 
$v_3 = \Lambda + \delta_3$. Let us denote $\delta_1-\delta_2$ and 
$\delta_2-\delta_3$ by $\Delta_{12}$ and $\Delta_{23}$, respectively. 
After taking  the limit of sending $\Lambda$ to infinity, 
we have 
\bea
A_{123}(\infty)& = & 1, \quad 
A_{132}(\infty)= {\frac {\Delta_{23} -2 i } {\Delta_{23} +2i}}, \quad 
A_{213}(\infty)= {\frac {\Delta_{12} -2 i } {\Delta_{12} +2i}},  \non \\
A_{231}(\infty) & =& \left( 
{\frac {\Delta_{12} -2 i } {\Delta_{12} +2i}}
\right) 
\left( 
{\frac {\Delta_{12} +\Delta_{23} -2 i } {\Delta_{12} +\Delta_{23} +2i}} \right) , \non \\ 
A_{312}(\infty) & = &
\left( 
{\frac {\Delta_{12} +\Delta_{23} -2 i } {\Delta_{12} +\Delta_{23} +2i}} \right) 
\left( 
{\frac {\Delta_{23} -2 i } {\Delta_{23} +2i}} \right) , \non \\
A_{321}(\infty) & = & 
\left( 
{\frac {\Delta_{12} -2 i } {\Delta_{12} +2i}} \right) 
\left( 
{\frac {\Delta_{12} +\Delta_{23} -2 i } {\Delta_{12} +\Delta_{23} +2i}} \right) 
\left( 
{\frac {\Delta_{23} -2 i } {\Delta_{23} +2i}} \right) . 
\label{samples3}
\eea
The limit of the Bethe ansatz wavefunction for  
the formal Bethe state $||0, 3; \Lambda \ra$ is given by 
\be 
\lim_{\Lambda \rightarrow \infty}
f_{0, 3}^{(B)} (x_1,x_2,x_3; k_1(\Lambda), k_2(\Lambda), k_3(\Lambda)) = 
C_3
\ee
where the constant $C_3$ is given by 
\bea
C_3 & = & \sum_{P \in S_3} A_3(P) \left[\delta_1,\delta_2,\delta_3 \right]
 \non \\
& = & 1 + {\frac {\Delta_{23} -2 i } {\Delta_{23} +2i}}+ 
{\frac {\Delta_{12} -2 i } {\Delta_{12} +2i}}
   + \left( 
{\frac {\Delta_{12} -2 i } {\Delta_{12} +2i}}
\right) 
\left( 
{\frac {\Delta_{12} +\Delta_{23} -2 i } {\Delta_{12} +\Delta_{23} +2i}} \right)  \non \\ 
 & + &
\left( 
{\frac {\Delta_{12} +\Delta_{23} -2 i } {\Delta_{12} +\Delta_{23} +2i}} \right) 
\left( 
{\frac {\Delta_{23} -2 i } {\Delta_{23} +2i}} \right)  +  
\left( 
{\frac {\Delta_{12} -2 i } {\Delta_{12} +2i}} \right) 
\left( 
{\frac {\Delta_{12} +\Delta_{23} -2 i } {\Delta_{12} +\Delta_{23} +2i}} \right) 
\left( 
{\frac {\Delta_{23} -2 i } {\Delta_{23} +2i}} \right) . \non \\
\eea
Thus, we have 
\bea
 \lim_{\Lambda \rightarrow \infty} ||0,3; \Lambda \ra  
& = & C_3 \sum_{1 \le x_1<x_2<x_3 \le L}
\sigma_{x_1}^- \sigma_{x_2}^{-} \sigma_{x_3}^{-} |0 \ra  \non \\
& = & C_3 \, {\frac 1 {3!}} \left( S_{tot}^{-} \right)^3 |0 \ra  \non \\
& = & C_3 \, |0,3 \ra 
\eea


\setcounter{equation}{0} 
\renewcommand{\theequation}{C.\arabic{equation}}
\section{Appendix:  Some useful properties of the  symmetric group}

\par 
We introduce some notation of the symmetric group  
\cite{Jacobson}. Let $M$ be  a positive integer. 
We  consider  the permutation group  $S_M$ of integers  $1, 2, \ldots, M$. 
Take an element $P$  of $S_M$.  
We denote the action of $P$ on $j$  by $Pj$ for $j =1, \ldots, M$. 
Let us introduce  the symbol $(i_1 i_2 \cdots i_r)$ 
of the cyclic permutation  
 where $i_j$ is sent to $i_{j+1}$ for
$j=1, \ldots, r-1$,  and $i_{r}$ is sent to $i_1$.  
It is known \cite{Jacobson}  
that any permutation 
$P$ can be decomposed into a product of 
{\it disjoint cycles} such as follows 
\be
P= (i_1 i_2 \cdots i_r) 
(j_1 j_2 \cdots j_s) \cdots 
(\ell_1 \ell_2 \cdots \ell_u) .   
\label{factor}
\ee
Here, any two of the cycles 
share no letter (or integer) in common. 
The factorization (\ref{factor}) is unique 
except for order of the factors \cite{Jacobson}. 

\par 
For a given permutation $P$ with a  factorization of disjoint cycles such as 
eq. (\ref{factor}),  
we denote by $N(P)$ the sum $(r-1) + (s-1) + \cdots + (u-1)$.  
Then, we can show that 
the parity of the permutation $P$ is equal to that of $N(P)$. 
Hereafter, we shall write by the symbol $ a \equiv b$ (mod 2) 
that integers $a$ and $b$ have the same parity.  
We first recall that the cycle $(i_1 i_2 \cdots i_r)$ can be written as 
the product of $r-1$ transposition such as
$$
(i_1 i_2 \cdots i_r) = (i_1 i_r)(i_1 i_{r-1}) \cdots (i_1 i_2).  
$$ 
Thus, the parity of the cycle is given by that of $r-1$. 
Let us denote by the symbol $\epsilon(P)$  
the sign of  permutation $P$. 
Then, we have \cite{Jacobson} 
\be 
\epsilon(P)= (-1)^{(r-1) + (s-1) + \cdots + (u-1)} = (-1)^{N(P)}
\label{signN}
\ee

\par 
Let us introduce  ordered pairs of integers. 
We take two different integers $j$ and $k$,  
and consider an ordered pair $<j,k>$. 
We distinguish  $<j,k>$ from $<k,j>$. 
Let us consider the action of a permutation  
on ordered pairs. We take a permutation  $P$ of $S_M$ 
and two integers $j$ and $k$ satisfying $1 \le j < k \le M$.  
We denote by $<Pj, Pk>$ the action of $P$ on the pair $<j,k>$.    
If $Pj > Pk$,  we call the pair $<j,k>$ is transposed by $P$.

\par 
Let the symbol  $T(P)$ denote  the number of all such pairs 
$<j,k>$ that are transposed by $P$  among all the 
ordered pairs $<j,k>$ with the condition $1 \le j < k \le M$. 
Then, we can show the following. 
\begin{lem}
The parity of an element $P$ of  $S_M$ 
is equivalent to that of the number $T(P)$: 
\be
N(P) \equiv T(P) \quad ({\rm mod} \, 2)  .
\label{mod}
\ee
\end{lem}
({\it Proof}) 
We now prove the lemma based on induction on $M$ of $S_M$. 
It is easy to see that when $M=2$ the statement is true. 
Let us now assume that 
eq. (\ref{mod}) holds for all permutations $P$ of 
$S_R$ if $R < M$. 
Let us take an element $P$ of $S_M$. 
Then, we may assume that the permutation $P$ has   
a factorization of disjoint cycles such as shown in eq. (\ref{factor}).  
Suppose that $P$ has the same factorization with eq. (\ref{factor}). 
We take a cycle $(i_1i_2 \cdots i_r)$, which is  one of 
the disjoint cycles, and we denote  by $B$ the set $\{i_1,i_2, \ldots, i_r  \}$. 
We also denote by $\Sigma_M$ the set of $M$ integers:    
$\Sigma_M = \{ 1, 2, \ldots, M \}$. 
We now consider  the subset $A$ of the set $\Sigma_M$ 
that is complementary to the set $B$: 
$A= \Sigma_M - B$.   
We define permutation $P_A$  by 
\be
P_A= 
(j_1 j_2 \cdots j_s) \cdots (\ell_1 \ell_2 \cdots \ell_u) . 
\ee 
Note that $P_A$ is a permutation of $A$ and 
it does not change any letter in $B$: 
$P_A i_j= i_j$ for $j=1,\ldots, r$. Thus,  we see that  
$T(P)$ and $T(P_A) + (r-1)$ have the same parity. 
Here, we note that $T((i_1i_2 \cdots i_r))
= r-1$, i.e., 
$r-1$ pairs of the elements in 
$B$ are transposed by $(i_1i_2 \cdots i_r)$,  
and therefore by $P$. 
On the other hand, since $P_A$ is  a permutation of $A$, 
it is equivalent to an element of $S_{M-r}$.  
From the induction hypothesis, we have that  
$N(P_A)$ and $T(P_A)$ have the same parity. 
Thus, we have 
\bea
T(P) & \equiv & T(P_A) + (r-1) \quad ({\rm mod} 2)  \non \\
& \equiv &  N(P_A) +(r-1) \quad ({\rm mod}  2)  \non \\
& = & N(P) \non
\eea
Therefore, $T(P)$ and $N(P)$ have the same parity.  
\begin{flushright}
{\it Q.E.D.} 
\end{flushright}

We now have the following . 
\begin{prop}
Let $P$ be an element of $S_M$.  
Then, we have the following identity. 
\end{prop}
\be
\epsilon(P) 
=
\prod_{1\le j<k \le M} 
(-1)^{H(P^{-1}j-P^{-1}k)}  
\label{transposed}
\ee
({\it Proof}) 
Let us  note the following  
\be 
T(P) = \sum_{1 \le j < k \le M} {H(P^{-1}j-P^{-1}k)}  
\ee
Then, we can show eq. (\ref{transposed}) 
 from the previous lemma and eq. (\ref{signN}).


\setcounter{equation}{0} 
\renewcommand{\theequation}{D.\arabic{equation}}
\section{Appendix: Proof of the ``Pauli principle''} 

We give a simple proof for the ``Pauli principle''  
of the Bethe ansatz  that when there are two rapidities 
of the same value, then the Bethe ansatz wavefunction 
of the XXX model  vanishes. 
We note that it is also proven by  the algebraic Bethe ansatz method 
in Ref. \cite{Korepin-book}. 
However, the proof in this appendix is much more elementary;  
it is only based on the expression (\ref{amp}) of the amplitudes $A_M(P)$'s. 
In this appendix, we assume that 
rapidities $v_1 \ldots, v_M$ are free parameters. 

\par 
Let us take a pair of integers $a$ and $b$ such that  
$1 \le a < b\le M$.   
Then, we show that the Bethe ansatz wavefunction $f_M^{(B)}$ 
with the amplitudes defined by eq. (\ref{AM1}) (equivalently by eq. (\ref{amp})) 
vanishes if $k_a = k_b$ ({\it i.e.},  $v_a = v_b$).  
Let the symbol  $(ab)$ denote the permutation between $a$ and $b$. 
Then,  we have 
\bea 
& & f^{(B)}_M(x_1, \ldots, x_M; k_1, \ldots, k_M) 
 =  \sum_{P \in {\cal S}_M} 
A_M(P) \exp(i\sum_{j=1}^{M}k_{Pj}x_j) \non \\
&= & 
{\frac 1 2} \sum_{P \in {\cal S}_M} A_M(P) 
\exp\left(i \sum_{j=1}^{M} k_{Pj} x_j \right)  
 + {\frac 1 2} \sum_{P \in {\cal S}_M} A((ab)P) 
\exp\left(i \sum_{j=1}^{M} k_{((ab)P)j} x_j \right)  
\eea
Here we have replaced $P$ by $(ab)P$ in the second term. 
Considering the cases when $j=P^{-1}a$ and $j=P^{-1}b$, 
we can show  
\bea 
\sum_{j=1}^{M} k_{Pj} x_j
& = & k_a x_{P^{-1}a} + k_b x_{P^{-1}b} +
\sum_{j=1; \, j \ne P^{-1}a, P^{-1}b }^{M} k_{Pj} x_j  \\ 
 \sum_{j=1}^{M} k_{((ab)P)j} x_j 
 & = & 
 k_{(ab)a} x_{P^{-1}a} + k_{(ab)b} x_{P^{-1}b} +
\sum_{j=1; \, j \ne P^{-1}a, P^{-1}b }^{M} k_{(ab)Pj} x_j
\non \\
& = & 
 k_{b} x_{P^{-1}a} + k_{a} x_{P^{-1}b} +
\sum_{j=1; \, j \ne P^{-1}a, P^{-1}b }^{M} k_{Pj} x_j
\eea
When $k_a = k_b =k $, we have 
\bea 
f^{(B)}_M(x_1, \ldots, x_M; k_1, \ldots, k_M) 
& = &
{\frac 1 2} \sum_{P \in {\cal S}_M} \Big( A_M(P) +A_M((ab)P))   \Big)
\non \\
& & \times \exp\left(ik(x_{P^{-1}a} +x_{P^{-1}b}) +  
i \sum_{j=1; \, j\ne P^{-1}a , P^{-1}b }^{M} k_{Pj} x_j \right)  
\eea

We now show that $A_M(P) + A_M((ab)P)=0$ for any $P \in {\cal S}_M$. 
Here we introduce the following symbols
\be 
e(j,k) = {\frac {v_j -v_k-2i} {v_j -v_k +2i}} \, ,   \qquad H(j,k;P) = H(P^{-1}j - P^{-1}k) 
\ee
Then, the  amplitude $A_M(P)$ given by eq.(\ref{amp}) is expressed as 
\be 
A_M(P) =  \prod_{1 \le j < k \le M} e(j,k)^{H(j,k; P)} 
\ee
Let us  consider the six cases for the integers $j$ and $k$ 
 in the above product: 
 $j=a$ and $k=b$;  $j < a$ and $k=a$;  $j < b$ and $k=b$ where $j \ne a$; 
 $j=b$ and $k > b$; $j = a $ and $k > a $ where $k \ne b$; $j \ne a$ and $k \ne b$. 
  We have the following   
\bea 
A_M(P) & = & e(a,b)^{H(a,b; P)} \, 
\prod_{j=1}^{a-1} e(j,a)^{H(j,a; P)}  \prod_{j=1; \, j \ne a}^{b-1} e(j,b)^{H(j,b; P)}  \non \\
& & \times \prod_{k=b+1}^{M} e(b,k)^{H(b,k; P)}  \prod_{k=a+1; \, k \ne b}^{M} e(a,k)^{H(a,k; P)} 
\prod_{1 \le j < k \le M; \, j,k \ne a,b} e(j,k)^{H(j,k; P)} \non \\ 
& = & e(a,a)^{H(a,b; P)} \, 
\prod_{j=1}^{a-1} e(j,a)^{H(j,a; P)+ H(j,a;P)}  \prod_{j=a+1}^{b-1} e(j,a)^{H(j,b; P)}  \non \\
& & \times \prod_{j=b+1}^{M} e(a,j)^{H(b,j; P)+H(a,j;P)} 
 \prod_{j=a+1}^{b-1} e(a,j)^{H(a,j; P)} \prod_{1 \le j < k \le M; \, j,k \ne a,b} e(j,k)^{H(j,k; P)} 
\non \\
& = & (-1)^{H(a,b; P)} \, \prod_{j=a+1}^{b-1} e(a,j)^{H(a,j; P)-H(j,b; P)} \times  
 \prod_{1 \le j < k \le M; \, j,k \ne a,b} e(j,k)^{H(j,k; P)} \non \\
& & \times \prod_{j=1}^{a-1} e(j,a)^{H(j,a; P)+ H(j,b;P)} 
           \prod_{j=b+1}^{M} e(a,j)^{H(b,j; P)+H(a,j;P)} 
  \eea
Here we have used the relations $e(j,a)=e(j,b)$, $e(a,j)=1/e(j,a)$, 
$e(a,b)=e(a,a)=-1$, and so on. 
In a similar way, we have 
\bea 
A_M((ab)P) & = & (-1)^{H(b,a; P)} \, \prod_{j=a+1}^{b-1} e(a,j)^{H(b,j; P)-H(j,a; P)}  
\times  \prod_{1 \le j < k \le M; \, j,k \ne a,b} e(j,k)^{H(j,k; P)}  \non \\
& & \times \prod_{j=1}^{a-1} e(j,a)^{H(j,a; P)+ H(j,b;P)} 
           \prod_{j=b+1}^{M} e(a,j)^{H(b,j; P)+H(a,j;P)} 
\eea
Noting the relation: $H(j,k; P)-1/2 = - (H(k,j; P) -1/2)$, we can show  
\bea 
H(a,j; P)-H(j,b; P) & = & H(P^{-1}a - P^{-1}j) - H(P^{-1}j - P^{-1}b) \non \\
                    & = & - H(-P^{-1}a + P^{-1}j) + H(-P^{-1}j + P^{-1}b) \non \\
                   & = & - H(j,a; P) + H(b,j; P) \, . 
\eea
Thus, we have 
\bea 
& & A_M(P) +A_M((ab)P) =  \left( (-1)^{H(a,b; P)}+ (-1)^{H(b,a; P)} \right) 
 \prod_{j=a+1}^{b-1} e(a,j)^{H(a,j; P)-H(j,b; P)}  \non \\
& & \times \prod_{j=1}^{a-1} e(j,a)^{H(j,a; P)+ H(j,b;P)} 
           \prod_{j=b+1}^{M} e(a,j)^{H(a,j; P)+H(b,j;P)} 
 \prod_{1 \le j < k \le M; \, j,k \ne a,b} e(j,k)^{H(j,k; P)} \, , 
\eea
and we obtain   
\be 
A_M(P)+ A_M((ab)P) =0 \quad {\rm for } \quad {\rm any} \quad P \in S_M. 
\ee
Here we note the following:  ${H(b,a; P)}=0$ when ${H(a,b; P)}=1$;  
 ${H(b,a; P)}=1$ when ${H(a,b; P)}=0$.

\par 
Following  the discussion in the appendix, 
we can show the Pauli principle of the Bethe ansatz also for the XXZ model;    
we redefine $e(j,k)$  by 
 $e(j,k) = {\sinh(v_j - v_k + 2 \eta)}/{\sinh(v_j - v_k - 2 \eta)}$,   
where $\eta$ is related to $\Delta$ in eq. (\ref{XXZ}) by 
$\Delta = \cosh 2\eta$.


\setcounter{equation}{0} 
\renewcommand{\theequation}{E.\arabic{equation}}
\section{Appendix: Rigorous derivation of the coordinates Bethe ansatz} 

\subsection{Secular equations of the XXZ model}

\par 
The coordinate Bethe ansatz was introduced 
by Bethe for the one-dimensional XXX model in Ref. \cite{Bethe}. 
In this appendix, we  derive rigorously some sets of 
sufficient conditions for a vector to be an eigenvector of 
of the XXZ model under the twisted boundary conditions. 
The derivation should be useful for discussing singular eigenvectors 
of the model such as shown in Refs. \cite{sl2loop,XXZincomplete,complete} . 
We note that when the twisting parameter is zero: $\Phi=0$, 
 the twisted boundary conditions reduces into the periodic 
boundary conditions,  
and also that when $\Delta=1$ the XXZ Hamiltonian (\ref{XXZ}) 
becomes the XXX Hamiltonian.  

\par 
Let us consider the action of the XXZ Hamiltonian ${\cal H}_{XXZ}$ (\ref{XXZ})
on any given vector. 
We recall that the symbol $|M)$ in eq. (\ref{|M)}) 
denotes a vector with $M$ down-spins  
where the amplitude $g(x_1, x_2, \ldots, x_M)$ is given by any function. 
Then,  the action of ${\cal H}_{XXZ}$  
on the vector $|M)$ can be calculated rigorously. 
The result is given in the following
\bea
& & \left( {\cal H}_{XXZ} -J \Delta (M- {\frac L 4}) \right) \, |M) 
 = - {\frac J 2}\sum_{1 \le x_1 < \cdots < x_M \le L} 
\Bigg\{ \sum_{j=1}^{M} \sum_{s=\pm 1} 
g(x_1, \ldots, \overbrace{x_j +s}^{j{\rm th}}, \ldots, x_M) \non \\
& & -\sum_{j=1}^{M-1} \delta_{x_j+1, x_{j+1}} 
\bigg( g(x_1,\ldots, \overbrace{x_j,x_j}^{j, j+1},\ldots, x_M)  
    + g(x_1,\ldots, \overbrace{x_j+1,x_j+1}^{j, j+1},\ldots, x_M)  
    -2 \Delta g(x_1,\ldots, x_M) \bigg)  \non \\
&& - \delta_{x_1,1} \delta_{x_M,L}\left(  
g(0,x_2, \ldots, x_M) + g(x_1,  \ldots, x_{M-1}, L+1) 
- 2 \Delta g(1,x_2, \ldots, x_{M-1}, L) \right) 
\Bigg\}  \times \prod_{k=1}^{M} \sigma_{x_k}^{-} |0 \ra  \non \\
&& - {\frac J 2} \sum_{1 < x_1 < \cdots < x_{M-1} < L} 
\Bigg\{ \left( - g(0, x_1, \ldots, x_{M-1}) + g(x_1, \ldots, x_{M-1}, L) \, e^{-i \Phi} \,  \right) \, 
 \sigma_{1}^{-} \prod_{k=1}^{M-1} \sigma_{x_k}^{-} \, |0 \ra \non \\
& & \quad + \left(  g(1, x_1, \ldots, x_{M-1}) - g(x_1, \ldots, x_{M-1}, L+1) \, e^{-i \Phi} \,  \right) \, 
 \sigma_{L}^{-} \prod_{k=1}^{M-1} \sigma_{x_k}^{-} \, |0 \ra \Bigg\} 
\label{action} 
\eea
Here, we have assumed the twisted boundary conditions for the
spin operators:  ${\sigma}_{L+1}^{\pm}  = e^{\pm i \Phi} {\sigma}_{1}^{\pm}; \sigma_{L+1}^{z} = \sigma_{1}^{z} $, 
while any boundary conditions have been assigned 
on the function $g(x_1, \ldots, x_M)$.

\par 
Let us discuss sufficient conditions 
for vector $|M)$ to  be an eigenvector of the XXZ Hamiltonian, explicitly.  
 Considering  the last part of eq. (\ref{action}) we have the 
 twisted  boundary conditions on the function $g(x_1, \cdots , x_M)$: 
\bea 
 g(0, x_1, \ldots, x_{M-1}) & = & g(x_1, \ldots, x_{M-1}, L) e^{- i \Phi} \quad 
{\rm for} \, \, 1 < x_1 < \cdots < x_{M-1} < L \, , 
\non \\
g(1, x_1, \ldots, x_{M-1}) & = & g(x_1, \ldots, x_{M-1}, L+1) e^{- i \Phi} \quad 
{\rm for} \, \, 1 < x_1 < \cdots < x_{M-1} < L \, . 
\label{PBCg}
\eea 
 Considering the second part of eq.(\ref{action}), we have the following conditions.
\bea 
& & g(x_1, \ldots, x_j, x_j, \ldots,x_M)  +  
 g(x_1, \ldots, x_j+1, x_j+1, \ldots, x_M)   \non \\
& & \qquad - 2 \Delta g(x_1, \ldots, x_j, x_j+1, \ldots, x_M) =0 \non \\ 
& & \qquad {\rm for} \quad 1 \le x_1 <  \cdots < x_j, \, 
x_j+1  < x_{j+2} <\cdots < x_M \le L \quad {\rm and} \quad x_{j+1}=x_j+1 \, ,  
 \label{eigen} 
\eea 
where $j$ is given by  $j=1, \ldots, M-1$. And we also have 
\bea
& & g(0, x_2, \ldots,x_{M-1}, L) + g(1, x_2,  \ldots, x_{M-1}, L+1) 
- 2 \Delta g(1, x_2, \ldots,  x_{M-1}, L) = 0 \non \\ 
& & \qquad {\rm for} \quad 1  < x_2 <  \cdots < x_{M-1}  < L \, . 
\label{eigenb}
\eea
Under the twisted  boundary conditions (\ref{PBCg}), 
the conditions (\ref{eigenb}) correspond to the special 
cases of the conditions (\ref{eigen}), where  
$j=M-1$ and $x_{M-1}=L$, or $j=1$ and $x_1=0$.

\par 
Let us now assume that the function $g(x_1, \ldots, x_M)$ is given by 
 $f(x_1, \ldots, x_M)$ defined by  a general 
 linear combination of the planewave-type solutions 
\be 
f(x_1 , \ldots, x_M) 
= \sum_{P \in {\cal S}_M} B(P) \exp(i \sum_{j=1}^M k_{Pj} x_j) \, . 
\label{bethetype} 
\ee
Here $k_1, \ldots, k_M$ are free parameters, ${\cal S}_M$ denotes the 
symmetric group on $M$ letters, and the amplitudes $B(P)$'s are arbitrary. 
The  amplitudes $B(P)$'s in (\ref{bethetype}) are $M!$ independent parameters, 
and we shall determine them 
 so that the function $f(x_1, \ldots, x_M)$ satisfies 
the conditions (\ref{PBCg}), (\ref{eigen}) and (\ref{eigenb}). 
The function $f(x_1, \ldots, x_M)$ has the following property 
\bea
& & \sum_{j=1}^{M}  
\left(f(x_1, \ldots, x_{j-1}, x_j-1, x_{j+1}, \ldots, x_M) 
+ f(x_1, \ldots, x_{j-1}, x_j+1, x_{j+1}, \ldots, x_M)\right) \non \\ 
&  & = \left(  \sum_{j=1}^{M} 2 \cos k_j \right) f(x_1, \ldots, x_M) \, ,  
\quad {\rm for } \quad 1 \le x_1 < \cdots < x_M \le L \, .  
\label{cosine} 
\eea
In fact, we see  explicitly  
\bea 
& &  \sum_{j=1}^{M}  
\left(f(x_1, \ldots, x_j-1, \ldots, x_M) + f(x_1, \ldots, x_j+1, \ldots,
x_M)\right) \non \\ 
& = & 
\sum_{j=1}^{M} \sum_{P \in {\cal S}_M} \bigg( 
B(P) \exp(i \sum_{\ell=1}^{M}k_{P\ell}x_{\ell} -i k_{Pj}) + 
B(P) \exp(i \sum_{\ell=1}^{M}k_{P\ell}x_{\ell} + i k_{Pj}) \bigg) \non \\
& =& \sum_{P \in {\cal S}_M} \sum_{j=1}^M B(P) 
\exp(i \sum_{\ell=1}^M k_{P\ell}x_{\ell} )
\left( \exp(-ik_{Pj}) + \exp( i k_{Pj}) \right) \non \\
& =&  \left( \sum_{j=1}^{M} 2 \cos k_j \right)  f(x_1, \ldots, x_M). 
\eea

\par 
Let us summarize the discussion given in the above. 
Assuming that the function $g(x_1, \ldots, x_M)$ is given by 
$f(x_1, \ldots, x_M)$ defined by eq. (\ref{bethetype}), 
the vector $|M)$ is an eigenfunction of the Hamiltonian 
if the conditions (\ref{PBCg}), 
(\ref{eigen}) and (\ref{eigenb}) are satisfied.

\par 
We now show that the conditions  (\ref{eigen}) 
are satisfied if  the following relations hold 
for the amplitudes $B(P)$'s: 
\bea 
& & B(Q) \left( 1+ \exp(ik_{Qj}+ ik_{Q(j+1)}) - 
2 \Delta \exp(i k _{Q(j+1)}) \right) \non \\
&+  & 
 B(Q \pi_j) \left( 1+ \exp(ik_{Qj}+ ik_{Q(j+1)}) 
-2 \Delta \exp(i k _{Qj}) \right) =0 \non \\ 
& & \quad {\rm for } \quad Q \in {\cal S}_M \quad 
{\rm and } \quad j=1, \ldots, M-1. \label{vanishB}
\eea
Here the symbol $\pi_j$ denotes the permutation of $j$ and $j+1$:
$\pi_j = (j, j+1)$ for $j=1, \ldots, M-1$.  
Explicitly we have 
\bea
& & f(x_1, \ldots, x_{j-1}, x_j, x_j, x_{j+2}, \ldots, x_M) +
f(x_1, \ldots, x_{j-1}, x_j+1, x_j+1, x_{j+2}, \ldots, x_M) \non \\ 
& & \quad - 2 \Delta f(x_1, \ldots, x_{j-1}, x_j, x_j+1, x_{j+2}, \ldots, x_M) \non \\
&= & \sum_{P \in {\cal S}_M} \bigg\{ 
B(P) \exp \left(\sum_{\ell=1; \, \ell \ne j, j+1}^{M} 
i k_{P\ell}x_{\ell} + i k_{Pj} x_j + i k_{P(j+1)} x_j \right)   \non \\
& & \quad + 
B(P) \exp  \left(\sum_{\ell=1; \, \ell \ne j, j+1}^{M} 
i k_{P\ell}x_{\ell} + i k_{Pj} (x_j +1) + i k_{P(j+1)} (x_j+1) \right)  
\non \\ 
& & \quad - 2  \Delta 
B(P) \exp  \left( \sum_{\ell=1; \, \ell \ne j, j+1}^{M} 
i k_{P\ell}x_{\ell} + i k_{Pj} x_j + i k_{P(j+1)} (x_j+1) \right)  \bigg\}  
\non \\
& = & 
\sum_{P \in {\cal S}_M} B(P) \exp  \left( \sum_{\ell=1; \, \ell \ne j, j+1}^{M} 
i k_{P\ell} x_{\ell} + i (k_{Pj} + k_{P(j+1)}) x_j \right)  \non \\
& & \quad \left( 1+ \exp(ik_{Pj}+ ik_{P(j+1)}) -2 \Delta \exp(i k _{P(j+1)}) \right) 
\non \\ 
&=  & {\frac 1 2}
 \sum_{Q \in {\cal S}_M}  \exp \left( \sum_{\ell=1; \, \ell \ne j, j+1}^{M} 
i k_{Q \ell} x_{\ell} + i (k_{Qj} + k_{Q(j+1)}) x_j \right) \non \\
& & \times \bigg\{ 
B(Q) \left( 1+ \exp(ik_{Qj}+ ik_{Q(j+1)}) -2 \Delta \exp(i k _{Q(j+1)}) \right) 
\non \\
& & +  B(Q \pi_j) 
\left( 1+ \exp(ik_{Qj}+ ik_{Q(j+1)}) -2 \Delta \exp(i k _{Qj}) \right) 
\bigg\} .
\label{QQ}
\eea 
Here, we have made use of the following relation
\be 
\sum_{P \in {\cal S}_M} F(P)= {\frac 1 2} \sum_{Q \in {\cal S}_M} F(Q) + 
{\frac 1 2} \sum_{Q \in {\cal S}_M} F(Q \pi_j) .  
\ee
There are $(M-1)M!/2$ independent 
relations in  eq. (\ref{vanishB}).  
If they hold, then the terms 
 involving  $B(Q)$'s  and $B(Q \pi_j)$'s  
in RHS of  eq. (\ref{QQ}) vanish. 
Thus,  LHS of (\ref{QQ}) becomes zero,   
and the conditions (\ref{eigen}) are satisfied for the case of $j$. 
Thus, we have shown that the relations (\ref{vanishB}) 
are sufficient for the conditions (\ref{eigen}). 
Hereafter, we shall call the conditions (\ref{vanishB}) 
the vanishing conditions.   

\par 
Let us discuss the number $W$ of independent relations given by 
eqs.  (\ref{eigen}) and (\ref{eigenb}). 
It is given by the number of configurations where $x_1, \ldots, x_M$ 
satisfy the conditions: $1  \le x_1 <  \cdots < x_j$, 
$x_j+1  < x_{j+2} <\cdots < x_M \le L$ and $x_{j+1}=x_j+1$ 
for $j=1, \ldots, M-1$ and also those of (\ref{eigenb}).  
The number $W$  is given by  
\be 
W = L \times  _{L-2}C_{M-2} = {\frac {L (L-2)!} {(L-M)! (M-2)!}} 
\ee
We recall that the number $V$ of the relations of (\ref{vanishB}) 
is given by 
$M!$. 

\par 
Let us now consider the ratio $W/V$. We recall that 
the vanishing conditions 
(\ref{vanishB}) are sufficient  for the conditions (\ref{eigen}) and  
(\ref{eigenb}) to hold. If the ratio $W/V$ is larger than 1, then 
the vanishing conditions (\ref{vanishB}) are also necessary conditions 
for (\ref{eigen}) and (\ref{eigenb}). Here we recall that the variables 
$k_1, \ldots, k_M$ are assumed to be  free  parameters in this subsection.  
Let us calculate the ratio $W/V$, explicitly. It is given by 
\be 
W/V = {\frac {2L (L-2)!} {(L-M)! M! (M-1)!} }
\ee
For example, we have $W/V=L$ for $M=2$; 
 $W/V=L(L-2)/6$ for $M=3$; $W/V=L(L-2)(L-3)/72$ for $M=4$. 
Thus,  when $L$ is large enough with respect to the number of down-spins $M$, 
then the ratio $W/V$ is larger than one. When $W/V > 1$, 
the vanishing conditions (\ref{vanishB}) 
are necessary for the relations (\ref{eigen}) and (\ref{eigenb}).  
However, we should note that the ratio $W/V$ is not always larger than 1. 
For instance, when $L=16$ and $M=8$, we have $W/V = 143/420 $. 
For the half-filling case, we have $W/V < 1$ for $M=L/2 \ge 8$, in general.   
If $W/V < 1 $, then the vanishing conditions (\ref{vanishB}) for the amplitudes 
are not necessary for the relations (\ref{eigen}) and (\ref{eigenb}).

\par 
Let us make a conclusion of Appendix E.A. 
Assuming that the function $g(x_1, \ldots, x_M)$ is given by 
$f(x_1, \ldots, x_M)$ defined by eq. (\ref{bethetype}),  the vector 
$|M)$ is an eigenvector of the XXZ Hamiltonian, 
if the vanishing conditions (\ref{vanishB}) and the twisted 
boundary conditions (\ref{PBCg}) hold.

\par 
We give remarks. Let us consider the case of $|\Delta|< 1$,  
where $\Delta = \cosh 2 \eta$.  
When all the momenta are generic, or 
$k_{Qj} \ne \pm 2 |\eta| \, ({\rm mod} \, 2 \pi )$ or 
$k_{Q(j+1)} \ne \pm 2 |\eta| \, ({\rm mod} \, 2 \pi)$, 
the relations (\ref{vanishB}) can be expressed  as follows
\bea
{\frac {B(Q)} {B(Q \pi_j)}} & = & (-1) 
{\frac  {1+ \exp(ik_{Qj} + ik_{Q(j+1)}) -2 \Delta \exp(ik_{Qj})}
{1+ \exp(ik_{Qj} + ik_{Q(j+1)}) -2 \Delta \exp(ik_{Q(j+1)})}}, \non \\
& & \quad {\rm for } \quad Q \in S_M \quad 
{\rm and } \quad j=1, \ldots, M-1. \label{factorB}
\eea
For the singular solutions of the  XXZ model such as  
discussed in Ref. \cite{sl2loop,XXZincomplete,complete},   
some of them satisfy the sufficient conditions 
(\ref{vanishB}) and (\ref{PBC}), but their amplitudes $B(P)$'s 
do not satisfy the factorization property (\ref{factorB}). 
 Some details will be discussed elsewhere.

\subsection{Amplitudes of the Bethe ansatz wavefunction}

We show that the amplitudes $A_M(P)$'s defined by eq. (\ref{AM1}) 
satisfy the vanishing conditions (\ref{vanishB}), where   
the parameters $k_1, \ldots, k_M$ are  generic. 
Let us introduce the following notation 
\be 
F_{j\ell}(P) =  \exp[i(k_{Pj}+k_{P{\ell}})] +1 -2 \Delta \exp(ik_{Pj})    
\ee
Then, the expression (\ref{AM1}) of the amplitude $A_M(P)$  is given by 
\be
  A_M(P) = 
   \epsilon(P) \prod_{1 \le j < \ell \le M} 
  {\frac 
   {F_{j \ell}(P) }
   {F_{j \ell}(e) } },  
  \quad {\rm for } \quad P \in {\cal S}_M . 
\label{AM1e}
\ee
Here $e$ denotes the unit element of the permutation group ${\cal S}_M$, 
and we put $C=1$ in (\ref{AM1}). 

\par 
Let us calculate the ratio of $A_M(Q \pi_a)$ and $A_M(Q)$, 
explicitly. Here $a $ is taken to be  
an integer satisfying $1 \le a \le M-1$. 
We note  that  integers  $j$ and $\ell$ 
satisfy the condition: $1 \le  j < \ell \le M$ in  (\ref{AM1e}). 
We consider the  four cases: (1) $j=a$ and $\ell=a+1$;   
 (2) $j=a$ and $\ell > a+1$;  (3) $j=a+1$ and $\ell > a+1$; 
(4) $j, \ell \ne a, a+1$. 
Then we have 
\be 
A_M(Q) \prod_{j < \ell} F_{j \ell}(e) = \epsilon(Q) F_{a a+1}(Q) 
\prod_{\ell > a+1} F_{a \ell}(Q) 
\prod_{\ell > a+1} F_{a+1 \ell}(Q)  
\prod_{1 \le j < \ell \le M; \, j,\ell \ne a, a+1}
F_{j \ell}(Q)
\ee
 When $P=Q\pi_a$, we have  
\bea 
A_M(Q \pi_a ) \prod_{j < \ell} F_{j \ell}(e) & = & \epsilon(Q\pi_a) 
F_{a a+1}(Q \pi_a) 
\prod_{\ell > a+1} F_{a \ell}(Q \pi_a) 
\prod_{\ell > a+1} F_{a+1 \ell}(Q \pi_a )  \non \\
& & \qquad \times 
\prod_{ j < \ell ; \, j,\ell \ne a, a+1}
F_{j \ell}(Q \pi_a)
\eea
Through an explicit calculation, we have 
\bea 
F_{a a+1}(Q \pi_a) & = &  F_{a+1, a}(Q) \, , \quad    
F_{a+1 a}(Q \pi_a)  =  F_{a, a+1}(Q) \non \\
F_{a \ell}(Q \pi_a) & = & F_{a+1, \ell}(Q) \, , \quad 
F_{a+1 \ell}(Q \pi_a)  =  F_{a, \ell}(Q)
\eea
Thus, the ratio is given by 
\be 
{\frac {A_M(Q)} {A_M(Q\pi_a)}}  = (-1) {\frac {F_{a, a+1}(Q)} { F_{a+1, a}(Q)}  }
\ee
It is nothing but the relation (\ref{factorB}), 
which is equivalent to 
the vanishing conditions (\ref{vanishB}) for the case of $a$. 
Here we recall that all the momenta are given generic in Appendix E.B. 

\par 
Let us now  discuss 
 the uniqueness or  well-definedness 
of the amplitudes $A_M(P)$'s.
First we note that  any permutation $P$ can be written 
in terms of a product of generators $\pi_j$'s. 
For example, cyclic permutation $(213)$ corresponds to  $(23)(12)=\pi_2 \pi_1$. 
Thus, we can calculate $A_M(P)$ by using the relations (\ref{factorB}). 
For example, let us take $P=(213)$. We have 
\bea 
 A(\pi_2 \pi_1) & = &  {\frac {A(\pi_2 \pi_1)}  {A(\pi_2 )}}
{\frac   {A(\pi_2)}  {A(e)}} A(e) \non \\
& = & {\frac {F_{21}(\pi_2)}{F_{12}(\pi_2)}}
       {\frac {F_{32}(e)}{F_{23}(e)}}
\eea
However,  different  products of generators 
can correspond to the same permutation. 
 For instance, we have  $(213) = (23)(12) = (12)(23)(12)(23)$. 
The amplitude $A(\pi_2 \pi_1)$ should be 
equivalent to  $A(\pi_1\pi_2 \pi_1\pi_2 )$.

\par
In fact, we can prove the uniqueness 
of amplitudes $A_M(P)$'s, explicitly.  
We note that  the defining relations 
\cite{Magnus} of the symmetric group 
${\cal S}_M$ given by the following:  
\bea
\pi_j^2 & = & 1  \quad {\rm for} \quad j=1, \ldots, M-1 \, , \label{defrel1} \\
\pi_j \pi_{j+1} \pi_j & = & \pi_{j+1} \pi_{j} \pi_{j+1} 
\quad {\rm for} \quad j=1, \ldots, M-2 
\label{defrel2}
\eea
Thus, for a given permutation $P$, 
any given two products of generators 
expressing the same $P$ can be transformed into one another, 
by using the defining relations given in eqs. (\ref{defrel1}) and (\ref{defrel2}). 
Therefore, the uniqueness of the amplitudes is proven  
if we show that the amplitudes $A_M(P)$'s satisfy  the following relations: 
\bea 
& & {\frac {A_M(Q)} {A_M(Q \pi_j)}}
\cdot 
{\frac {A_M(Q\pi_j)} {A_M((Q \pi_j) \pi_j)}}  =  1 
\quad {\rm for }\quad j=1, \ldots, M-1 \, ,  
\label{con1b} \\ 
& & {\frac {A_M(Q)} {A_M(Q \pi_j)}}
\cdot 
{\frac {A_M(Q\pi_j)} {A_M((Q \pi_j) \pi_{j+1})}} 
\cdot 
{\frac {A_M(Q\pi_j \pi_{j+1})} 
{A_M(Q \pi_j \pi_{j+1} \pi_j)}} \non \\
& = & 
{\frac {A_M(Q)} {A_M(Q \pi_{j+1})}} \cdot 
{\frac {A_M(Q\pi_{j+1})} {A_M((Q \pi_{j+1}) \pi_{j})}} 
\cdot 
{\frac {A_M(Q \pi_{j+1} \pi_j )} {A_M((Q \pi_{j+1} \pi_j) \pi_{j+1})}} . 
\quad {\rm for } \quad j=1, \ldots, M-2 \, .
 \label{con2b}  
\eea 
Here $Q \in {\cal S}_M$. 
In fact, it is easy to check the relations (\ref{con1b}) and (\ref{con2b}). 

\subsection{Derivation of the Bethe ansatz equations}

The  twisted boundary conditions (\ref{PBCg}) for 
the wavefunction are given by the following  
\be
f(x_1,\ldots, x_M) = \exp(-i \Phi)  f(x_2, \ldots, x_M, x_1 + L)  \quad 
{\rm for} \quad 0 \le x_1 < \cdots  < x_M \le L 
\label{PBC}
\ee
In terms of the amplitudes $B(P)$'s,  
RHS of (\ref{PBC}) is given by 
\bea 
& & 
\sum_{P \in {\cal S}_M} B(P) \exp\left(-i \Phi +i(k_{P1}x_2 + 
\cdots k_{P(M-1)}x_M + k_{PM}(x_1+N)) \right) \non \\
 & = & 
\sum_{P \in {\cal S}_M} B(P) \exp\left(-i \Phi +i k_{PM} N + i(
k_{PM} x_1 + k_{P1} x_2 + 
\cdots k_{P(M-1)}x_M ) \right) \non \\
 & = & 
\sum_{Q \in {\cal S}_M} B(Q(12 \cdots M)) 
\exp\left(-i \Phi + i k_{Q1} N + i \sum_{j=1}^{M}k_{Qj} x_j 
\right) . 
\label{RHS}
\eea 
Here we note that for $Q(12\cdots M) =P$ we have 
\be 
Q1=PM, \quad Q2=P1,\ldots, \quad QM=P(M-1) . 
\ee
In order for 
 RHS (\ref{RHS}) to be equivalent to  LHS of (\ref{PBC}), we have 
\be 
\exp(i k_{Q1} N ) = \exp(i \Phi) \, {\frac {B(Q)} {B(Q(12 \cdots M))} }, 
\quad {\rm for} \quad Q \in S_M .  
\label{primitive}
\ee

\par 
Let us now derive the Bethe ansatz equations 
from the eqs. (\ref{primitive}). Here, 
we assume that the amplitudes $B(P)$'s are given by 
 $A_M(P)$'s satisfying  the vanishing conditions (\ref{vanishB})
(or (\ref{factorB})).  
Then , we have  the following 
\bea 
{\frac {A(Q)} {A(Q(12 \cdots M))} } & = & 
{\frac {A(Q) } {A(Q\pi_{1})}} \cdot 
{\frac {A(\pi_{1}) } {A(Q\pi_{1} \pi_{2})}}
\cdots 
{\frac {A_M(Q\pi_{1}\cdots \pi_{M-2} ) } 
{A_M(\left( \pi_{1} \cdots \pi_{M-2} \right) \pi_{M-1})}} \non \\
& =& (-1)^{M-1} \prod_{m=2}^{M} \left( 
{\frac {1 + \exp(i(k_{Q1}+ k_{Qm}) - 2 \Delta \exp(ik_{Q1})}
{1 + \exp(i(k_{Q1}+ k_{Qm})) - 2 \Delta \exp(ik_{Qm})} }
\right)  \non \\
\eea
Here we note that $Q \pi_1 \cdots \pi_{r-1} r = Q1$ and 
$Q \pi_1 \cdots \pi_{r-1} (r+1) = Q(r+1)$. 
Thus, we have
\bea
\exp(iLk_{Q1}) &= & (-1)^{M-1} \exp(i \Phi) 
\prod_{m=2}^{M} \left( 
{\frac {1 + \exp(i(k_{Q1}+ k_{Qm}) - 2 \Delta \exp(ik_{Q1})}
{1 + \exp(i(k_{Q1}+ k_{Qm})) - 2 \Delta \exp(ik_{Qm})} }
\right) \non \\
& & \quad {\rm for}  \quad  Q \in S_M . 
\label{BAinQ} 
\eea
Let us write $Q1$ by $j$. Then, $Q2, \ldots, QM$ are given by 
all the integers from 1 to $M$ except $j$. We may write 
$Qm$ by $\ell$ which runs from 1 to $M$ except $j$. 
Thus, we obtain the standard form of the 
Bethe ansatz equations.

\par 
Finally, we summarize the rigorous formulation 
 of the coordinate Bethe ansatz. 
Assuming that the function $g(x_1, \ldots, x_M)$ is given by 
$f(x_1, \ldots, x_M)$ defined by eq. (\ref{bethetype}), 
we have shown in Appendix E.A that the vector 
$|M)$ is an eigenvector of the XXZ Hamiltonian, 
if the vanishing conditions (\ref{vanishB}) and the twisted  
boundary conditions (\ref{PBCg}) hold. 
In Appendix E.B, we have shown that the amplitudes 
$A_M(P)$'s defined by eq. (\ref{AM1}) are well-defined 
and also that they  indeed 
satisfy the vanishing conditions (\ref{vanishB}). 
In Appendix E.C, 
we have shown that the twisted  boundary conditions (\ref{PBCg}) 
are satisfied when the Bethe ansatz equations (\ref{BAinQ}) hold. 
Therefore, if $k_1, \ldots, k_M$ satisfy the Bethe ansatz equations, 
with the amplitudes $A_M(P)$'s constructed  by the momenta 
via eq. (\ref{AM1}), the vector $|M)$ becomes an eigenvector 
of the XXZ model under the twisted boundary conditions.


%


\end{document}